\documentclass[prd,tightenlines,showpacs,letterpaper, preprintnumbers,nofootinbib,floatfix]{revtex4}

\usepackage{amsmath,amssymb,amsfonts,graphicx,pifont,dcolumn}
\usepackage{epstopdf}
\usepackage{color}
\usepackage{cancel}
\usepackage{ulem}
\usepackage{xfrac}
\RequirePackage{slashed}

\begin{document}

\title{Looking for monopolium excited states with electromagnetic detectors}

\author{Huner Fanchiotti}  \author{C.A.~Garc\'\i a Canal}
\affiliation{ IFLP(CONICET) and Department of Physics, University of La Plata,  C.C. 67 1900, La Plata, Argentina}

\author{Marco~Traini} 

\affiliation{Dipartimento di Fisica, Universit\`a degli Studi di Trento, Via Sommarive 14, I-38123 Povo (Trento), Italy}

\author{ Vicente Vento }

\affiliation{Departamento de F\'{\i}sica Te\'orica and IFIC, Universidad de Valencia - CSIC, E-46100 Burjassot (Valencia), Spain.}

\date{\today}

\begin{abstract}
We study the detection of particles with magnetic moment and no charge by using coils and solenoids. The detector uses the Faraday-Lenz law to create a current whose energy is measured. Our development can be applied to any particle with magnetic moment. The neutron is ideal to show our detector scheme for particles with long lifetimes and the result is encouraging. However, our interest lies in the detection of monopolium states which decays much faster than neutrons. The accelerator scenario with a relatively low monopole mass $\sim 1000$ GeV requires the generalization of the scheme to finite lifetimes. We use two descriptions, which we have called Dirac and Schwinger, (also $\beta$-coupling), schemes, in our study. The theoretical  behavior of excited monopolia in coils teaches us many of their properties. However, their small production cross sections, given the achievable LHC luminosities, makes detection using coils difficult. Therefore, we generalize the discussion to cosmological heavy monopolium. Our scheme becomes useful for those and supplements detectors like MoEDAL and MAPP in distinguishing monopolia from other long lived particles.
\end{abstract}

\pacs{04.20.Jb, 04.50+h, 04.50Gh,04.80-y}

\maketitle

\section{Introduction}

Monopoles and their detection has been a matter of much research since Dirac discovered his now famous quantization condition~\cite{Dirac:1931qs,Dirac:1948um},

\begin{equation}
e g =\frac{N}{2} \;\;\; N= 1,2,\ldots
\end{equation}
where $e$ is the electron charge, $g$ the monopole magnetic charge and we have used natural units $\hbar=c=4\pi \varepsilon_0=1$. 

In Dirac's formulation, monopoles are assumed to exist as point-like particles and quantum mechanical consistency conditions lead to establish the magnitude of their magnetic charge. Monopole physics took a dramatic turn when 't Hooft~\cite{tHooft:1974kcl} and Polyakov~\cite{Polyakov:1974ek} independently discovered that the SO(3) Georgi-Glashow model~\cite{Georgi:1972cj} inevitably contains monopole solutions. These topological monopoles are impossible to create in particle collisions because of their huge GUT mass~\cite{Preskill:1984gd} or in other models with lower mass because of their complicated multi-particle structure~\cite{Drukier:1981fq}. 

Since the magnetic charge is conserved, monopoles will be produced predominantly in monopole-antimonopole pairs. 
Inspired by the old idea of Dirac and Zeldovich~\cite{Dirac:1931qs,Zeldovich:1978wj} that monopoles are not seen free because they are confined by their strong magnetic forces we have studied a monopole-antimonopole bound state that we have called monopolium~\cite{Vento:2007vy}.

The discovery of monopole and dipole solutions in Kaluza Klein theories \cite{Kaluza:1921tu,Klein:1926tv} made these theories very exciting
from a theoretical point of view \cite{Gross:1983hb}. In particular the dipole solution, which is classically stable and therefore very long lived, forms a very
interesting state which we have also called monopolium in analogy with its decaying cousin in  gauge
theories. The Kaluza Klein monopolium is extremely massive with a mass of the order of the Planck mass and therefore impossible to produce in laboratories. However, there might be clouds of them in the cosmos which might enter our detectors~\cite{Vento:2020vsq}.

Much experimental research has been carried out into the search for monopoles \cite{Cabrera:1982gz,Milton:2006cp,MoEDAL:2014ttp,Acharya:2014nyr,Patrizii:2015uea}. However, our interest here lies in the detection of monopolium, who is charge-less but which has a magnetic moment in the presence of magnetic fields or in excited deformed states \cite{Vento:2020vsq,Vento:2019auh}.  This is a phenomenologically oriented theoretical investigation and therefore the study of the behavior of monopolium in the presence of coils serves to determine interesting properties and energy regimes of excited monopolia which may serve to guide more sophisticated experiments given the progress of electromagnetic technologies at present.

In what follows we are going to discuss methods to detect magnetic moments and how they might be used in the detection of monopolium. In Section \ref{faraday}  we describe the effect of a magnetic moment traversing a coil. In Section \ref{Sdetector} we construct a detector for particles with magnetic moment and no charges. In Section \ref{neutrons} we will show some estimates analyzing the effect of a neutron beam on the detector. In Section \ref{monopolium} we study the application to  monopolium excited states  in accelerator and cosmological physics. Finally in Section \ref{conclusion} we collect the conclusions of our study.
 
 \section{Effects due to a particle with magnetic moment  passing through a conducting coil}
\label{faraday}
Our subject of study is a particle which has a permanent or an induced magnetic dipole moment $\overrightarrow{\cal{M}}$. We call this particle dipole for short.  Assuming the size of the particle small compared with the distance at which we are measuring the magnetic field its dipole moment will produce a magnetic field \cite{Vento:2019auh}

\begin{equation}
\overrightarrow{B}_d (\vec{r})= \frac{ 3\overrightarrow{\cal{M}}\cdot \vec{r} \,\vec{r} - \overrightarrow{\cal{M}}\, r^2}{r^5},
\label{Bfield}
\end{equation}
where $r=|\vec{r}|= \sqrt{(x-x_0)^2 +(y-y_0)^2 +(z-z_0)^2}$, $(x,y,z)$ is any point in space and $(x_0, y_0, z_0)$ is the position of the dipole considered point-like for this calculation.

\begin{figure}[htb]
\begin{center}
\includegraphics[scale= 0.35]{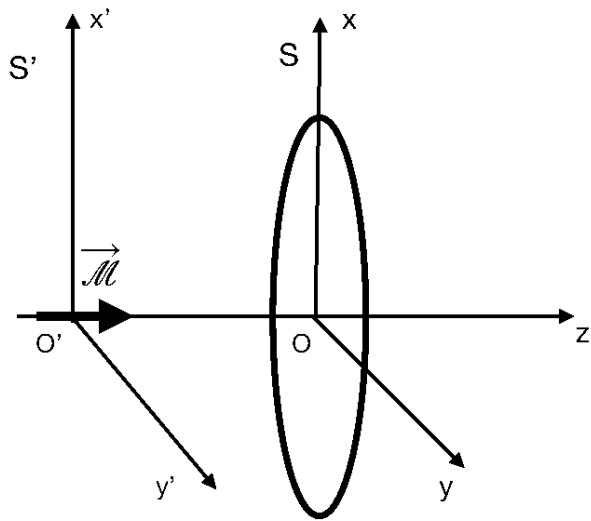} 
\end{center}
\caption{We show our set up: a particle traveling with constant velocity  towards a conducting coil with its magnetic dipole moment perpendicular to the plane of the coil. $S^\prime$ is the reference system associated to the dipole and $S$ the one associated to the coil.}
\label{coil}
\end{figure}

Let the dipole travel with a constant velocity towards a circular conducting coil of radius $R$ with its magnetic moment perpendicular to the plane of the coil as shown in Fig. \ref{coil}. Let $S^\prime$ be the reference system associated to the dipole and $S$ the reference system associated to the coil. We chose the coil to be located in  the $z=0$ plane, thus at time $t$ the dipole will be located at $z= - v t$.
The magnetic field felt by any point of space $\vec{r^\prime} = (x^\prime, y^\prime, z^\prime)$ when the dipole is  $\overrightarrow{\cal{M}} = {\cal M}\, \hat{k^\prime}$ is

\begin{equation}
\overrightarrow{B^\prime}_d (\vec{r^\prime})= \frac{3 \mathcal{M} \hat{k^\prime} \cdot \vec{r^\prime} \,\vec{r^\prime} -  {\mathcal M} \, \hat{k^\prime} \, r^{\prime 2}}{r^{\prime 5}},
\label{Bfield0}
\end{equation}
where  $r^\prime  = |\vec{r^\prime}|= \sqrt{x^{\prime 2} + y^{\prime 2} +z^{\prime 2}}$ and $\hat{k^\prime}$ is the unit vector in the $z$ direction. We need the magnetic field at the coil. The Lorentz transformation from system  $S^\prime$ reference system to the $S$ reference system is given by,

\begin{align*}
x^\prime = & x, \\
y^\prime = & y,\\
z^\prime = & \gamma (z -v t), \\
t^\prime =& \gamma(t - z),\\
\end{align*}
where $\gamma = 1/\sqrt{1- v^2}$. Since we have no electric field present the magnetic field changes \cite{Apyan:2007wq}

\begin{align}
B^\prime_x &  = \gamma B_x,\\
B^\prime_y & = \gamma B_y,\\
B^\prime_z & = B_z. 
\end{align}

With this geometry the flux through the coil is

\begin{equation}
\Phi(t) = \int^R_0 d \rho \int^{2 \pi}_0 d \varphi \,\overrightarrow{B}_d (x , y, - \gamma v t) \cdot \hat{k},
\label{flux}
\end{equation}
where $\rho$ and $\phi$ are the polar coordinates in the $z=0$ plane of the coil and $R$ is the radius of the coil. This calculation can be performed analytically and leads to

\begin{equation}
\Phi(t) = -\frac{2 {\cal M} \pi R^2}{((\gamma v t)^2 +R^2)^{3/2}}.
\label{flux1}
\end{equation}
\begin{figure}[htb]
\begin{center}
\includegraphics[scale= 0.9]{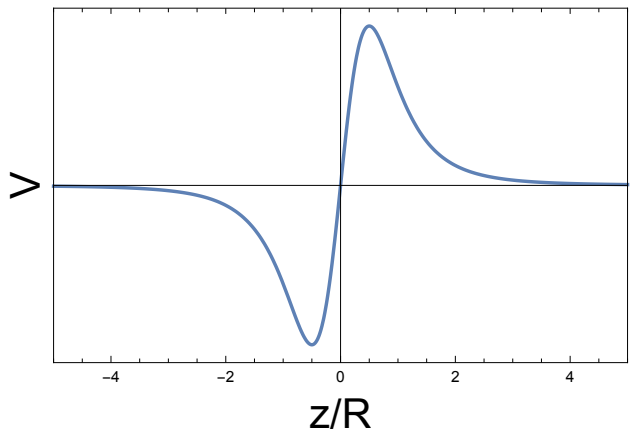} 
\end{center}
\caption{Functional form of the voltage generated by a dipole with a magnetic moment  passing a conducting coil at uniform speed.}
\label{emffig}
\end{figure}

Applying the Faraday-Lenz law the induced Electro Motive Force (EMF) generated by the magnetic field becomes

\begin{equation}
V(\gamma\, v\,t,R) =  \frac{ 6 {\cal M} \pi R^2 (\gamma v )^2 t}{((\gamma v t)^2 +R^2)^{5/2}}.
\label{emf}
\end{equation}
This equation can be written using $z(t) =  \gamma v t$, the area of the coil $A=\pi R^2$ and the strength parameter $\chi = 6 {\mathcal M} \gamma v$ as

\begin{equation}
V(z, A,\chi) =  \frac{\chi A z }{(z^2 + \frac{A}{\pi})^{\sfrac{5}{2}}}.
\label{emf1}
\end{equation}
This function can be seen in Fig.\ref{emffig} for fixed $A$ and $\chi$.

We have assumed in the previous calculation that the dipole passes through the axis of the coil. Ref. \cite{Apyan:2007wq} shows the result when the dipole does not pass through the axis, which is non analytic. Off axis the shape of the pulse is asymmetric and the time span shorter. For the purposes of the estimates which will be calculated in this presentation Eq.(\ref{emf1}) will be sufficient, with the advantage of being analytic.

A simple mathematical calculation determines the position of the extrema at

\begin{equation}
z=\pm \frac{1}{2} \sqrt{\frac{A}{\pi} }= \pm \frac{1}{2} R,
\label{extrema}
\end{equation}
where the voltage takes the value

\begin{equation}
V(\pm \frac{1}{2} \sqrt{\frac{A}{\pi} },A, \chi) = \pm \frac{16 \pi^2 \chi}{ 25 \sqrt{5} A}.
\label{Vextrema}
\end{equation}

The signal, a rise in the voltage, is governed by two parameters the strength $\chi$ and the area of the coil, being proportional to the former and inversely proportional to the latter, thus the ideal radius of the coil $R$ has to be as small as possible compatible with all the approximations.

\begin{figure}[htb]
\begin{center}
\includegraphics[scale= 0.85]{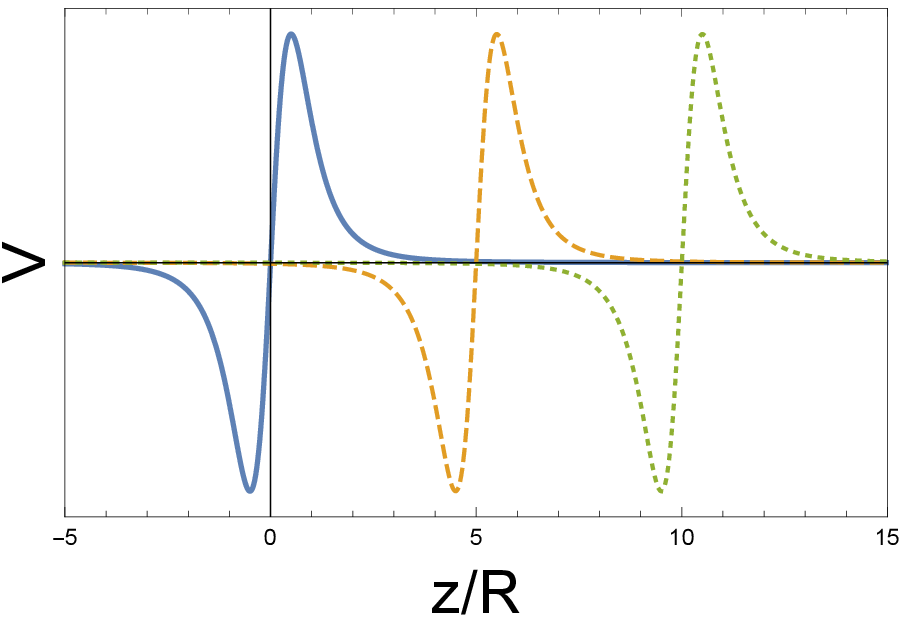} \hspace{1cm} \includegraphics[scale= 0.85]{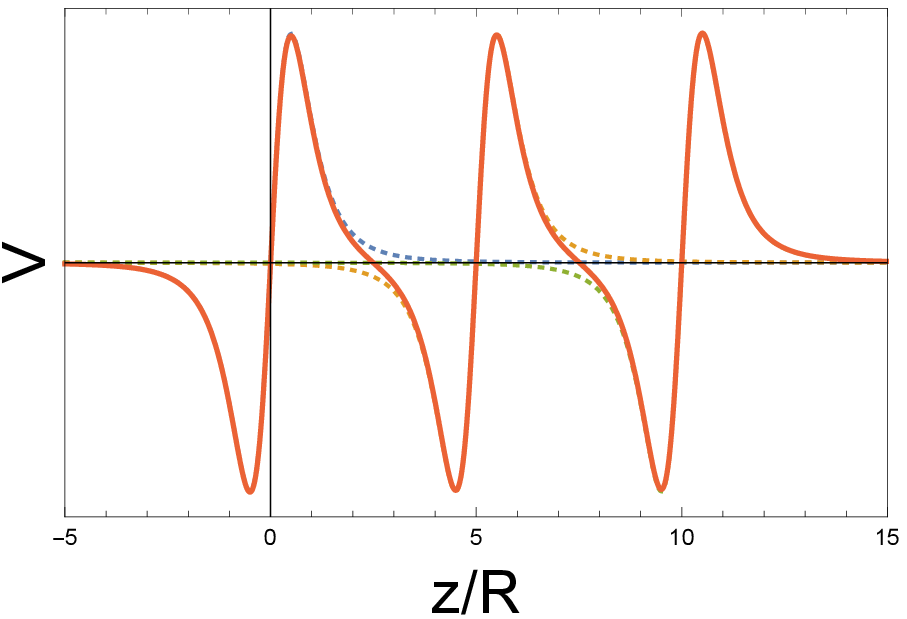} 
\end{center}
\caption{The left figure shows the functional form of the voltage generated by a particle with a magnetic moment when passing three independent coils at uniform speed. The figure on the right shows the combined effect of the three well separated coils forming a solenoid (solid curve) compared with the independent effect of the three coils (dotted curve). The effect is basically the same effect as that of the three coils independently.}
\label{separatedcoils}
\end{figure}

Let us introduce a system with a large number of coils close to each other forming a cylinder. There are two ways of creating this solenoid, either with well separated coils or with closely winded coils. In the first case, as shown in Fig. \ref{separatedcoils},  the effect is simply to add  the effect of each coil almost independently.

\begin{figure}[htb]
\begin{center}
\includegraphics[scale= 0.85]{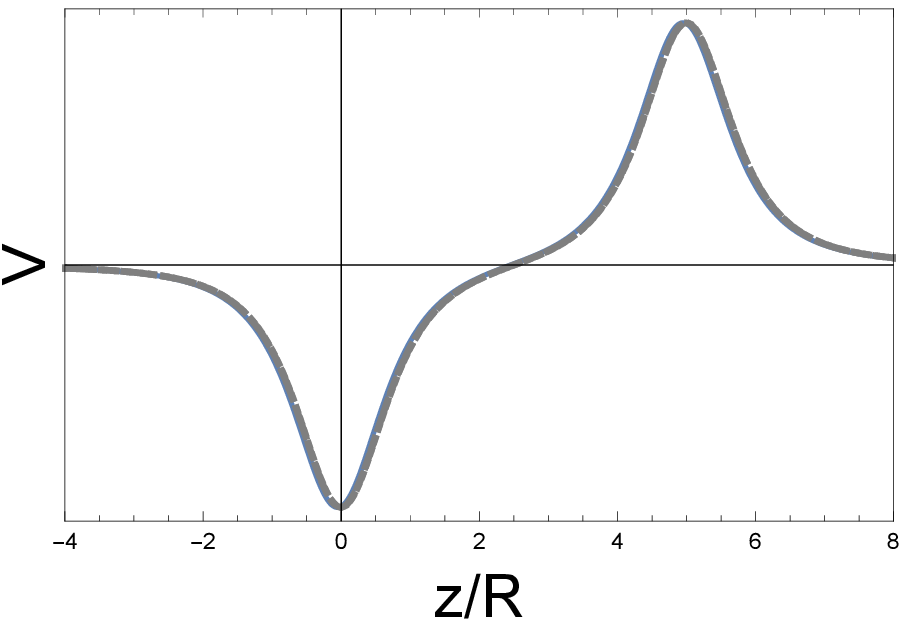} \hspace{1cm} \includegraphics[scale= 0.85]{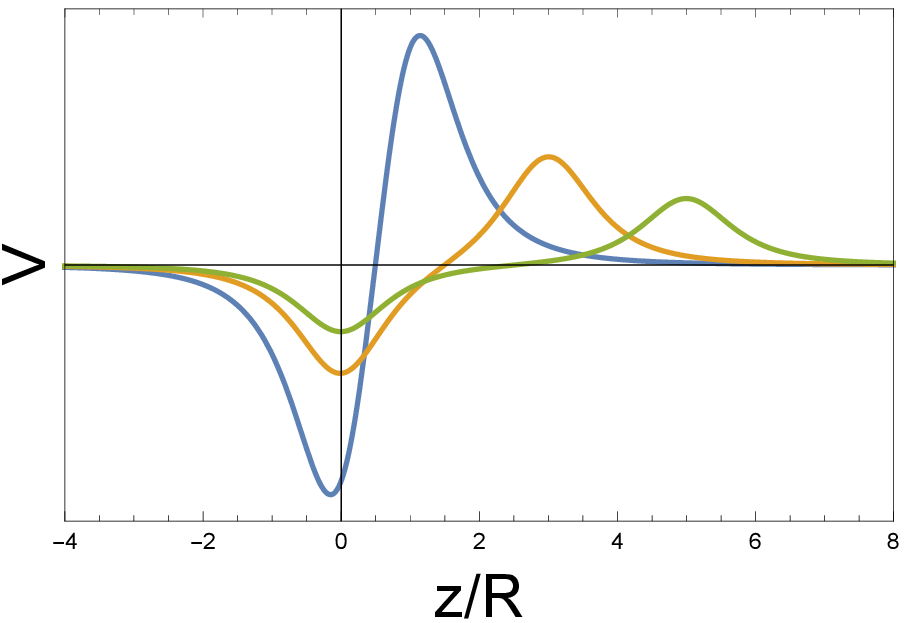} 
\end{center}
\caption{The left figure shows the functional form of the voltage generated by a particle with a magnetic moment when passing a conducting coils' system of length $5 R$  at uniform speed. The distance between the peaks is the length of the coils' system. The figure is composed by the calculation of the voltage for $50, 500$ and $5000$ loops divided by the number of loops showing that the proportionality between number of loops and voltage persists for relatively large lengths of the coils' system. The figure on the right shows the same for three different lengths of the coils' system. The shorter the length the greater the voltage.}
\label{faradayN}
\end{figure}

If the coils are very close the effect is more sophisticated the advantage being that we can put many coils in a shorter length to produce an enhancement of the coil effect in shorter distance. This case requires a detailed calculation that can be performed  numerically:

\begin{equation}
V_{N} (z, A, \chi, L) = \sum_{n=1}^{n=N} V(z - (n-1) \Delta z,A, \chi),
\label{exact}
\end{equation}
where $V$ is the potential in Eq.(\ref{emf1}), $N$ is the number of coils, $L$ is the length of the coil cylinder and $\Delta z =\frac{L}{N}$.
We show in  Fig. \ref{faradayN} the voltage divided by the number of coils generated by $50, 500$ and $5000$ coils located between $0\le\frac{z}{R}\le5$. This result is only valid for a large number of coils, when there is a cancellation between the positive and negative potentials. 

\begin{figure}[htb]
\begin{center}
\includegraphics[scale= 0.9]{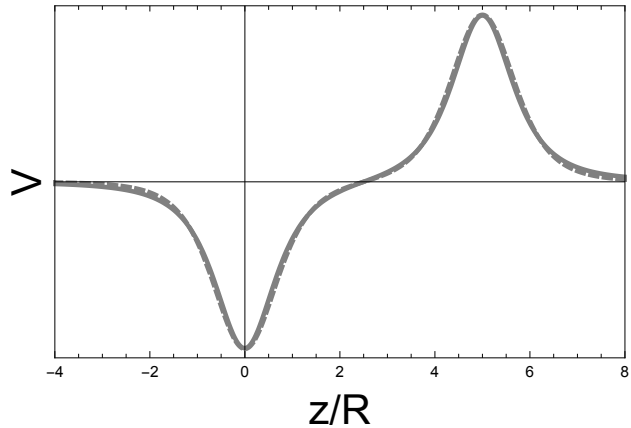} 
\end{center}
\caption{We show the functional form of the exact numerical calculation Eq.(\ref{exact})(solid) and the approximate one Eq.(\ref{solenoid})(dashed) for $N=1000$, $L = 5 R$  $A=0.1 R^2 $ and $\chi = 0.5$ V$\cdot R$.}  
\label{approximation}
\end{figure}

Doing some numerical analysis we find the relation between the maxima of the single coil potential and the $N$ coil potential for $N$ large

\begin{equation}
\frac{V_{N} (max)}{N} \sim 0.1312  \sqrt{A} \,V_1(max) \sim 0.3706 \frac{\chi}{\sqrt{A}}.
\end{equation}
Moreover, we are able to find an approximate analytical solution valid for large $N$

\begin{align}
\frac{V_{N \, approx}^{in}(z)}{N} &\sim \frac{ -b}{(z^2 + \frac{a}{\pi})^{\sfrac{5}{2}}}, \nonumber\\
\frac{V_{N \, approx}^{out} (z)}{N}& \sim \frac{ b}{((z - L)^2 + \frac{a}{\pi})^{\sfrac{5}{2}}}, \nonumber \\
&
\label{solenoid}
\end{align}
where $a = 2 A$ and $ b= 0.1198\, \chi \,A^2$. The $in$ potential describes the approximation to the  incoming minimum and the $out$ potential the approximation to the outcoming maximum. In Fig.\ref{approximation} we show the result of the exact calculation for $N=1000$ and compare it to  the approximation for the mentioned values of the parameters. 

\begin{figure}[htb]
\begin{center}
\hspace{-0.5cm}\includegraphics[scale= 0.13]{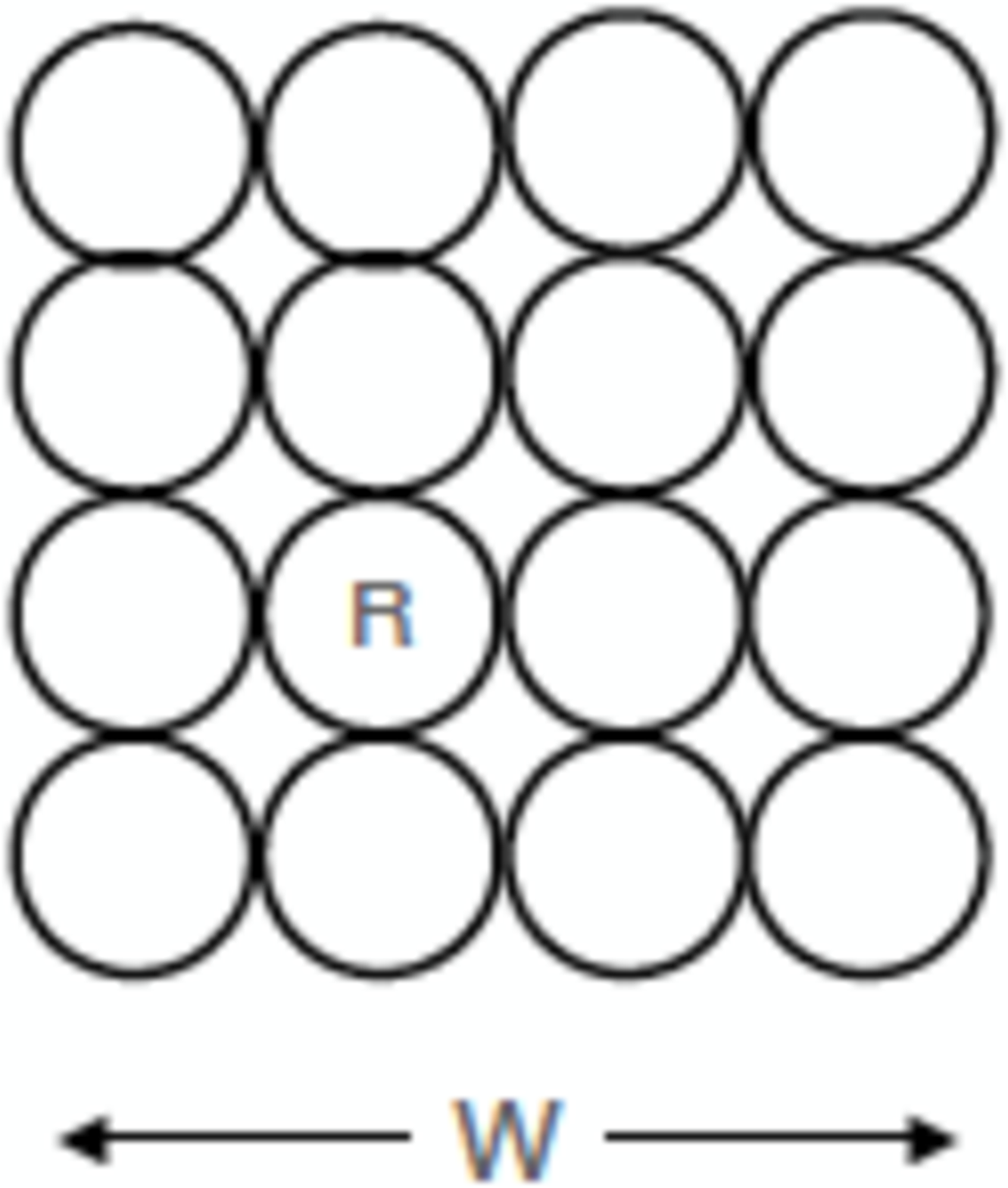} \hspace{-1.8cm} \includegraphics[scale= 0.13]{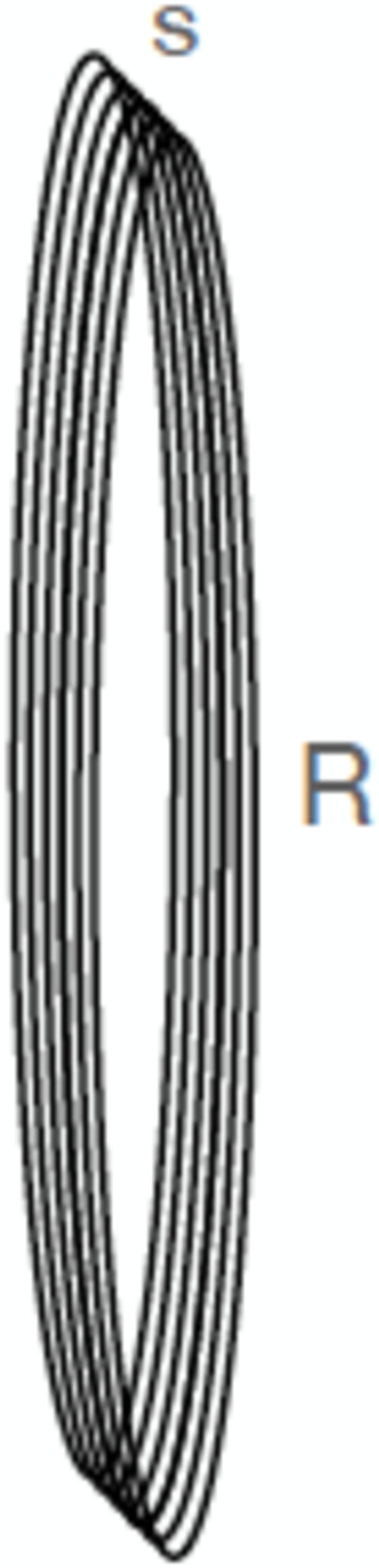} 
\end{center}
\caption{The left figure shows the ensemble of coils of a possible detector system where $W >>R$. The right figure shows the structure of a series of closely located coils $R>>s$}
\label{core}
\end{figure}

It is interesting to note from the above expression that the peak of the potential depends directly on the magnitude of the dipole moment and is inversely proportional to the square root of the area of the coil. This means that one does not need large coil sizes to construct a detector. Certainly there are limitations to the sizes due to the approximations used, i.e for circular coils $A= \pi R^2$  and $R$ has to be bigger than the size of the system. Thus the ideal detector to see such dipole in the scenario presented here seems to be the one shown  in Fig. \ref{core}, a detector with a large number of solenoids,  each formed by a large number of  coils,  all linked in a circuit in series to sum up the voltages. For a single particle the signal will be the voltage structure shown in Fig. \ref{faradayN} multiplied by the number of coils, where the separation between the voltage peaks is the length of the coil system. The result of the calculation shows that the voltage scales nicely with the number of coils and that the smaller the coil system the higher the potential for the same number of coils. Once has thus to find a compromise between length and number of coils. The width of the coil has not been taken into account in this calculation, but its effect is irrelevant at this point as long as its size $s$ is smaller than $R$.

\section{Determination of observables characterizing  particles with magnetic moment and no electromagnetic charges}
\label{Sdetector}

In the first subsection we will calculate the current created in systems of coils by the moving dipole as a means of characterizing particles with magnetic moment and no charge, while in the second subsection we will comment on the small effect on the particle motion of this induced current.

\subsection{The induced current and the deposited energy in coil systems}

A dipole traversing a coil  creates a magnetic field due to the ${\overrightarrow{\cal M}}\cdot \hat{k}$ component of its magnetic dipole moment whose flux  in the coil generates, via the Faraday-Lenz law, an EMF Eq.(\ref{emf}),

\begin{equation}
V_1 (z) = 6 {\mathcal M}_z \, \gamma \,v \, A\,z  \frac{1}{(z^2 + \frac{A}{\pi})^{\sfrac{5}{2}}}.  \\
\label{emf2}
\end{equation}

Only the $z$ component of the magnetic moment  $ {\overrightarrow{\cal M}} \cdot \hat{k}  = {\cal M}_z = {\cal M}\cos{\theta}$, where $\cos{\theta} =\frac{{\overrightarrow{\cal M}} \cdot \hat{k}}{{\cal M}}$ is active. Since we expect the dipoles initially in random directions  in the forward direction, the average ${\cal M}_z$ will be {\cal M} multiplied by the factor, 

\begin{equation}
\frac{\int_0^{2\pi} d\varphi \int_{-\frac{\pi}{2}}^{\frac{+\pi}{2}} d \theta \cos{\theta}}{\int_0^{2\pi} d\varphi \int_{-\frac{\pi}{2}}^{\frac{+\pi}{2}} d \theta } = \frac{2}{\pi}.
\end{equation}
and therefore Eq.(\ref{emf2}) becomes

\begin{equation}
V _1(z) = \frac{12}{\pi} {\mathcal M}\, \gamma \,v \, A\,z  \frac{1}{(z^2 + \frac{A}{\pi})^{\sfrac{5}{2}}} .\\
\label{coil1}
\end{equation}
where $ {\mathcal M}$ is the magnetic moment of the dipole.

For the $N$ coil solenoid, recalling Eq.(\ref{solenoid}), the EMF becomes

\begin{align}
\hspace{2.2cm}V_{N }^{in\;} (z) &\sim -0.4576\, {\mathcal M} \, \gamma \,v \, N\, A^2  \frac{1}{(z^2 + \frac{2 A}{\pi})^{\sfrac{5}{2}}}, \nonumber \\
V_{N }^{out } (z)& \sim +0.4576\, {\mathcal M} \, \gamma \,v \,N\, A^2 \frac{ 1}{((z - L)^2 + \frac{2 A}{\pi})^{\sfrac{5}{2}}},  \nonumber\\
&
\label{solenoid1}
\end{align}
where $N$ is the number of coils in each coil system, $L$ its length, $A$ is the area of the coil, $v$ the incoming velocity of the particle, $\gamma$ the Lorentz factor and ${\mathcal M}$ is the magnetic moment.

The total EMF has to take into account the inductance, thus

\begin{equation}
V_T =  - \frac{ d \Phi}{dt} - {\mathfrak L} \frac{d {\mathcal I}}{d t},
\end{equation}
where ${\mathfrak L}$ is the inductance and ${\mathcal I}$ the intensity of the circuit. Ohm's law then becomes the differential equation we have to solve in order to obtain the intensity in the circuit

\begin{equation}
\frac{ d \Phi}{dt} + {\mathfrak L} \frac{d {\mathcal I}}{d t} + {\mathcal I } {\mathcal R} = 0.
\end{equation}
where  ${\mathcal R} = \frac{2 \pi R}{S} \varrho N$, where $S$ is the surface of the conductor and $\varrho$ the resistivity which changes from conventional conductors, $\varrho= 10^{-8}$ Ohm$\cdot$m,  to superconductors $\varrho= 10^{-26}$ Ohm$\cdot$m. 

If $\mathcal R $ corresponds to a conductor the inductance term is small and only the induced EMF enters the equation, thus the intensity becomes ${\mathcal I }(z)  = - \frac{1}{\mathcal R} \frac{d \Phi}{dt} = \frac{V(z)}{\mathcal R} $ and therefore from Eqs.(\ref{coil}) and (\ref{solenoid1}) we get for one coil

\begin{equation}
{\mathcal I}_1(z) = \frac{12}{\pi} \frac{{\mathcal M}\, \gamma \,v \, A\,z }{ {\mathcal R}}\frac{1}{(z^2 + \frac{A}{\pi})^{\sfrac{5}{2}}} = \frac{12}{\pi} \frac{{\mathcal M}\, \gamma \,v \, A\,z }{ {\mathcal R}}\frac{1}{(z^2 + R^2)^{\sfrac{5}{2}}}, 
\label{coil2}
\end{equation}
and for the solenoid

\begin{align}
\hspace{1.9cm}{\mathcal I }_N^{in\,}(z)& = - 0.4576 \, \frac{{\cal M}\, \gamma  \,v \, A^2 \, N}{{\mathcal R} } \frac{1}{(z^2 + 2 R^2)^{\sfrac{5}{2}}}, \nonumber\\ 
{\mathcal I }_N^{out}(z)& = + 0.4576 \,\frac{ {\cal M}\, \gamma  \,v \, A^2 \, N}{{\mathcal R}}  \frac{ 1}{((z - L)^2 + 2 R^2)^{\sfrac{5}{2}}}. \nonumber \\
 &
 \label{solenoid2}
 \end{align}
If $ \mathcal R $ corresponds to a superconductor then the term ${\mathcal I}{ \mathcal R}$ is negligible and the intensity is given by ${\mathcal I} (z) = -\frac{\Phi}{{\mathfrak L}}  =  \frac{1} {\gamma\, v\, {\mathfrak L}} \int_{-\infty}^z V (z) dz  $, which becomes for the single coil

\begin{equation}
{\mathcal I}_1 = \frac{4}{\pi} \frac{{\mathcal M} A}{\mathfrak L} \frac{1} {(z^2 +\frac{A}{\pi})^{\sfrac{3}{2}}} = \frac{4}{\pi} \frac{{\mathcal M} A}{\mathfrak L} \frac{1} {(z^2 +R^2)^{\sfrac{3}{2}}},
\label{coil3}
\end{equation}
and for the $N$ coil solenoid

\begin{align}
\hspace{0.8cm}{\mathcal I}_N^{in} (z) & = -\frac{0.4576}{{\mathfrak L}} \, {\mathcal M}\, A^2 \, N \int_{-\infty}^z \frac{1}{(z^2 + \frac{2 A}{\pi})^{\sfrac{5}{2}}} dz   \nonumber \\ &= - \frac{0.3762}{{\mathfrak L}} \, {\mathcal M} \, N \, \frac{z(3 R^2 +z^2)}{(z^2 + 2 R^2)^{\sfrac{3}{2}} }, \nonumber\\
{\mathcal I}_N^{out} (z) & = \frac{0.4576}{{\mathfrak L}} \, {\mathcal M}\, A^2 \, N \int_{-\infty}^z \frac{1}{((z-L)^2+  \frac{2 A}{\pi})^{\sfrac{5}{2}}} dz \nonumber \\&= \frac{0.3762}{{\mathfrak L}} \, {\mathcal M} \, N \frac{(z-L)(3 R^2 +(z-L)^2)}{((z-L)^2 + 2 R^2)^{\sfrac{3}{2}}}. \nonumber \\
&
\label{solenoid3}
\end{align}

Let us study first the case of conducting coils, i.e. ${\mathcal R} \ne 0$ for stable particles\footnote{The case of decaying particles will be studied later on.}. The energy our detector can extract from the flux of dipoles is the integral of the power in time

\begin{equation}
{\cal E}  =  \int_{-\infty}^{+\infty} {\frac{V^2(t)}{{\cal R}} dt }.
\end{equation}
Thus for the single coil the integral is immediate and we get

\begin{equation}
{\mathcal E} = \left(\frac{12}{\pi}\right)^2 \frac{{\mathcal M}^2 \gamma v \, A^2 }{{\mathcal R}} \int_{-\infty}^{+\infty} \frac{z^2}{(z^2 + \frac{ A}{\pi})^5} dz = \frac{45}{8\pi} \frac{ {\mathcal M}^2 \gamma v }{R^3 {\mathcal R}}.
\label{energycoil}
\end{equation} 
For the solenoid the integral expression is

\begin{equation}
{\mathcal E} \sim 0.2094 \frac{ {\cal M}^2 \, \gamma \,v \, A^4\, N^2}{{\cal R}} \int_{-\infty}^{+\infty} \left(\frac{1}{(z^2 + \frac{2 A}{\pi})^{\sfrac{5}{2}} }+ \frac{1}{((z - L)^2 + \frac{2 A}{\pi})^{\sfrac{5}{2}}} \right)^2 dz,
\label{enerysolenoid1}
\end{equation}
and its integration requires some discussion.

The integral of the square terms can be performed analytically while the cross term not. In any case all of the three terms are very convergent and numerically we can get the result quite fast. However, the square terms tend to be much larger than the cross term, specially if $L$ is large compared to the width of the potential bumps, since the cross term only takes into account the overlap between the bumps. This argument has been checked numerically for the scenarios of the present calculation and therefore we neglect the cross term and get an analytic result which is 
\begin{eqnarray}
{\mathcal E} &\sim & 0.2094 \frac{ {\mathcal M}^2\,\gamma \,v \, A^4\, N^2}{{\mathcal R}} \frac{35 \pi}{64}\left(\frac{2 A}{\pi}\right)^{-\sfrac{9}{2}} \nonumber \\
&\sim & 2.7451 \frac{ {\mathcal M}^2\,\gamma \,v \, N^2}{{\mathcal R} \sqrt{A} }.
\end{eqnarray}
If we use the following units ${\cal M} $ (fm), $v$ in units of the velocity of light in the vacuum, $S$ (mm$^2$), $R$ (mm) and $\varrho$ (Ohm$\cdot$m) the equation for the energy becomes for the single coil

\begin{equation}
{\mathcal E} = 0.1682 \,10^{-29}\, \frac{ {\mathcal M}^2 \gamma v \, S }{R^4 \varrho } \, \mbox{eV}
\label{energycoil1}
\end{equation}
and for the solenoid

\begin{equation}
\hspace{0.5cm}{\mathcal E} = 0.5130 \, 10^{-20}\, \frac{ {\mathcal M}^2\, \gamma v\, S\, N}{R^2\, \varrho} \, \mbox{ eV}.
\label{normal}
\end{equation}

Let us now study the case of superconducting coils where the inductance term is dominant. In this case

\begin{equation}
{\mathcal E}  = \frac{ {\mathfrak L} }{2} \int_{-\infty}^{+\infty} \frac{ d{ \mathcal I}_N^2(t)}{dt } dt  = \frac{ {\mathfrak L}}{2} \, ( ({\mathcal I}_N^{in})^2(+\infty) + ({\mathcal I}_N^{out})^2(+\infty) )
 =  0.1415\, \frac{{\mathcal M}^2\,N^2}{{\mathfrak L}}
\end{equation}
For a thin long solenoid the ratio of inductance for superconducting and normal solenoids is close to 1 \cite{HIRAKAWA1973287}, thus  $ {\mathfrak L} = \mu_0 \frac{N^2 A}{L}$ where $ {\mathfrak L}$ is in H, if we take $\mu_0 = 4 \pi 10^{-7}$ H/m , $A$ the area of the coil m$^2$, $L$ the length of the coil in m. Recall that H = Ohm$\cdot$s, then

\begin{equation}
{\mathcal E} = 0.2486 \, 10^{-16} \, \frac{{\mathcal M}^2 L}{R^2} \, \mbox{eV}.
\label{super}
\end{equation}
where we are measuring ${\mathcal M}$ in fm, $R$ in mm and $L$ in m. This choice of three units might seem strange at first, but it is related to the natural scale of the physics and the apparatus: the magnetic moments of the states of monopolium are microscopic (0-1000 fm), the most effective coils are of mm size, and in order to have very large number of coils we need solenoids lengths of meters.

The behavior of Eq.(\ref{normal}) and Eq.(\ref{super})  are very different. For the former the parameters at our disposal to increase the sensibility of the detector are the number of coils  $N$, the velocity of the dipole $v$, but specially the factor $\gamma$, the radius of the coil $R$ and the resistivity of the material $\varrho$. For the latter the length of the solenoid $L$ and the radius of the coil $R$ but neither the velocity nor the number of coils play a role.

\subsection{Effect of the induced current on the velocity of the particle}

One might wonder if the induced current can produce a force which decelerates considerably the initial velocity of the particle. For simplicity we will perform a non relativistic calculation which is very transparent. Let us assume a particle traveling with constant velocity  towards a conducting coil with its magnetic dipole moment perpendicular to the plane of the coil as shown in Fig. \ref{coil}. We would like to find out the motion of this particle due to the effect of the Faraday-Lenz law which creates a magnetic field $\vec{B} = B_z \hat{k}$ where

\begin{equation}
B_z = \frac{2 \pi R^2 {\mathcal I}}{(z^2 +R^2)^{\sfrac{3}{2}}},
\end{equation}
where ${\mathcal I}$ is the intensity running through the coil, $z$ is the distance from the particle to the center of the coil and $R$ is the coil radius. Newtons's equation for the problem is given by

\begin{equation}
M \frac{d^2z}{dt^2}= - {\mathcal M} \frac{d B_z}{dz}.
\end{equation}

\begin{figure}[htb]
\begin{center}
\includegraphics[scale= 0.8]{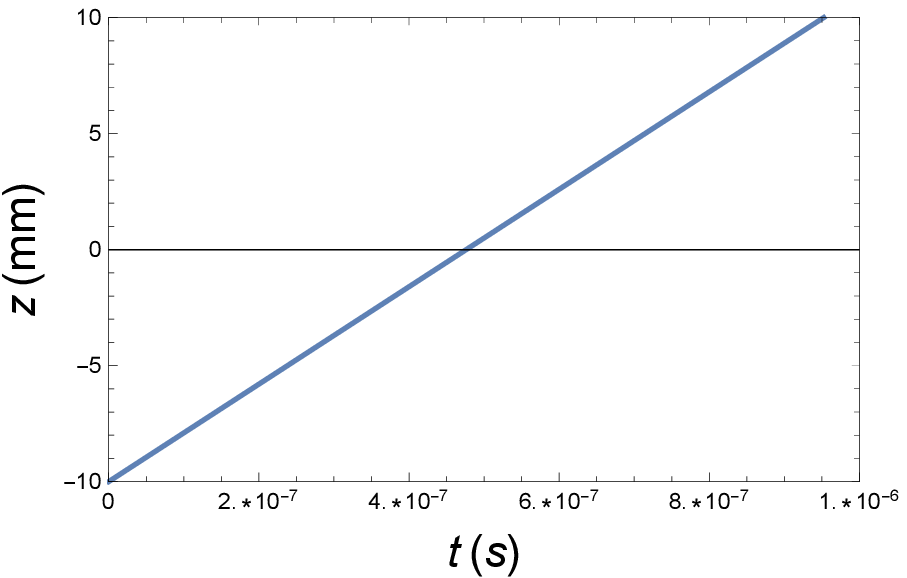} \hspace{0.5cm}\includegraphics[scale= 0.77]{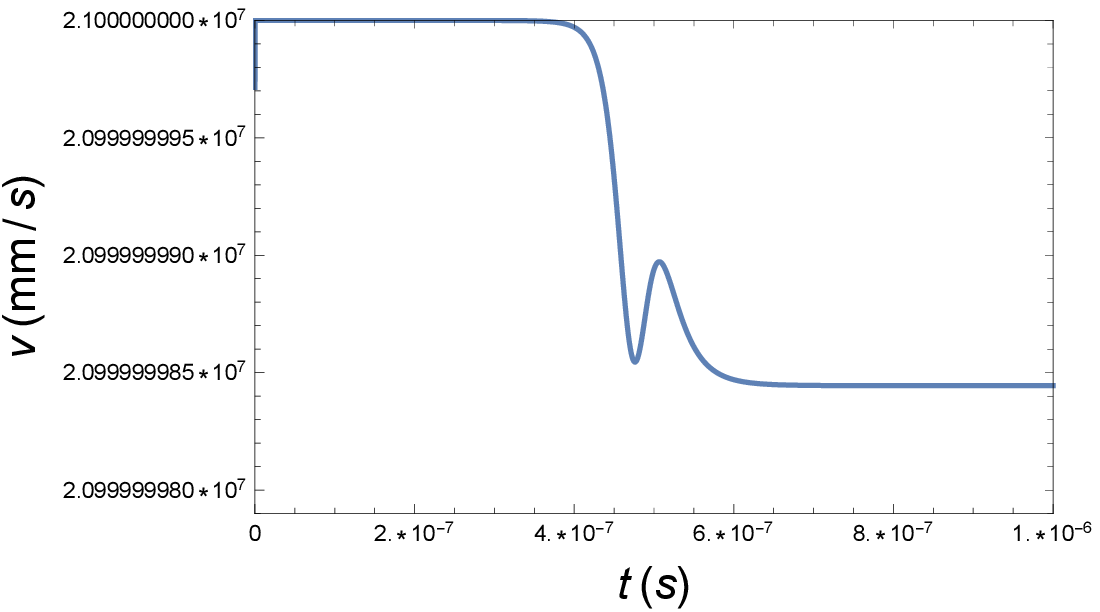} 
\end{center} 
\caption{A high excited monopolium (${\mathcal M} \sim 300 $ fm) traveling  with constant velocity ($v=7\;10^{-5}$) from $z_i= -10$mm towards a conducting coil located at $z=0$ with its magnetic dipole moment perpendicular to the plane of the coil.The left figure shows the variation of the position with time and the right figure the change in velocity with time.}
\label{FuerzaEspiraEx}
\end{figure}

In order to get the intensity we use Ohm's law

\begin{equation}
\frac{ d \Phi}{dt} + {\mathcal L} \frac{d {\mathcal I}}{d t} + {\mathcal I } {\mathcal R} = 0.
\label{ohm}
\end{equation}
where ${\mathcal L}$ is the inductance of the coil, ${\mathcal R}$ the resistance and the magnetic flux $\Phi$ is given by

\begin{equation}
\Phi(t) = -\frac{2 {\cal M} \pi R^2}{(z^2 +R^2)^{3/2}}.
\label{flux1}
\end{equation}
We are dealing with a very good conductor,
${\mathcal R} = \frac{2 \pi R}{\pi r_s^2} \varrho $, where $r_s$ is the radius of the conductor and $\varrho$ the resistivity  $\varrho= 10^{-8}$ Ohm$\cdot$m and has a small  inductance ${\mathcal L} =0.1$ nH. The set of equations that determine the motion of the particle are

\begin{eqnarray}
M \frac{d^2 z}{d t^2} & = & - 6 \pi {\mathcal M} R^2 {\mathcal I} \frac{z}{(z^2 +R^2)^{\sfrac{3}{2}} },\\
 {\mathcal L} \frac{d {\mathcal I}}{d t} + {\mathcal I } {\mathcal R} &= & 6 \pi {\mathcal M} R^2 {\mathcal I} \frac{z}{(z^2 +R^2)^{\sfrac{3}{2}} }\frac{d z}{dt}.
\end{eqnarray}
This is a system of differential equations for $z(t)$ and ${\mathcal I}$ (t) which has to be solved simultaneously. The initial conditions will be $z_i ,v_i$ and ${\mathcal I}_i$. We shall take $z_i$ sufficiently far from the coil so that the induced current is very small initially. The parameters that influence the solution  are ${\mathcal L}, {\mathcal R}, {\mathcal M}$ and $v_i$. The outcome of our calculation is shown in Fig. \ref{FuerzaEspiraEx}.

As can be seen in in Fig. \ref{FuerzaEspiraEx}, the particle approaches the coil at large distances without changing its velocity, Close to the coil, for large ${\mathcal M}$, the  coefficient of the derivative becomes sizeable, generating an acceleration in the opposite direction of the motion, which reduces the velocity. The velocity thus decreases steadily as the monopolium traverses the coil but only in its proximity. However, the effect on the velocity is very small as can be seen  in the figure, where we have used $z_i=-10 $mm, $v_i= 7.\,10^{-5}$, $t= 10^{-6}$s, and a very high lying Rydberg state with  ${\mathcal M} \sim 300$ fm.  It is interesting to note the small oscillation in the velocity at the origin which is an effect of the interplay between inductance and resistance. Thus the induced current does not influence significantly the calculation and we do not take it into consideration in what follows.

\section{Application to conventional particle detection: neutrons}
\label{neutrons}

Let us apply our formalism to a well known particle, the neutron.  We start using a solenoid made of a conventional conductor, $\varrho \sim 10^-8$ Ohm$\cdot$m. The simplest way to get a flux of neutrons is from a reactor from which one can obtain a beam of thermal neutrons with a velocity $v \sim 2000 m/s$.  In the case of the neutron ${\mathcal M}$ is given in natural units by

\begin{equation}
{\mathcal M} \sim -1.91\frac{e \hbar}{2 m_p} \sim - 6.1 \,10^{-2} \, \mbox{ fm}.
\end{equation}
Note that we have used in the presentation the classical approximation $\frac{2}{\pi}$ which is very close to the quantum mechanical $\frac{m}{s(s+1)}=\frac{1/2}{3/4} =2/3$ for the neutron.

 Thus our experiment consists of a beam of neutrons hitting a solenoid. Thus since $v \sim 2000$ m/s $\sim 0.67 \, 10^{-5} c$, $\gamma$ is very close to $1$. We take a thousand coils ($N=1000$ ) of ten millimeter radius ($R = 10\;mm$) and a surface of the conductor of $S= \pi r_s^2$ and $r_s =1$ mm, then the accumulated energy for one neutron is 

\begin{equation}
{\mathcal E} \sim 0.4018\; 10^{-18} \, \mbox{eV}.
\end{equation}

We send a beam of thermal neutrons from a reactor with a flux typically of $10^{13}$  neutrons  cm$^{-2}\,$ s$^{-1}$ to the detector. Let  the coil system has a surface covered by $10$ solenoids, as in Fig. \ref{core} right, all connected in series. In $1$ s the energy accumulated will be 
 
 \begin{equation}
 { \mathcal E} (1s) \sim 0.4018\, 10^{-4}  \,\mbox{eV}.
 \end{equation}
In a few days of  running one accumulates ${\mathcal E} \sim$ eV which is easily measurable. Since the velocity of the beam is known, ${\mathcal M}$ becomes the only unknown. Therefore the coil system could serve as a detector to determine the magnetic moment of the neutron. 

If we use a superconductor solenoid assuming the solenoids of $L=1$ m length,

 \begin{equation}
 { \mathcal E} (1s) \sim 0.9250\, 10^{-8}  \,\mbox{eV},
 \end{equation}
the difference arises because there is no $N$ dependence in the superconducting case, since the inductance grows also with $N^2$. A superconducting material is not favored in this case. 

For the proposed detector, even for a single coil, superconductors are not favored over normal conductors, thus we will use the latter from now on.

\section{Application to unconventional particle detection: monopolium}
\label{monopolium}

We discuss two types of experimental scenarios:  one related to a relatively light monopolium ($M \sim 2000$ GeV) which can be created by particle collisions~\cite{Epele:2012jn,Mavromatos:2020gwk} and one related to a heavy monopolium ($ M\sim M_{GUT}$ and $M \sim M_{Planck}$) that might be created in cosmological scenarios~\cite{Preskill:1984gd,Vento:2020vsq}.  

\subsection{Accelerator monopolium}
\label{acceleratormonopolium}

Let us suppose that we have to measure a  flux of light monopolia using their magnetic moment. This case resembles the neutron case except by the fact that we have considered the neutron as a stable state due to its long lifetime and monopolia might decay faster, and therefore their lifetime has to be taken into account. Let us implement this feature  in our description. Monopolium  is produced in particle collisions with small cross sections \cite{Epele:2012jn,Mavromatos:2020gwk}. To describe the momopole-antimonopole potential in monopolium there are several models in the literature \cite{Goebel1970,PhysRev.160.1257,Barrie:2016wxf,Baines:2018ltl,Barrie:2021rqa}. For the purpose of our present investigation the approximation to the potential of Schiff and Goebel \cite{Goebel1970,PhysRev.160.1257} 

\begin{equation}
V(r) = -g^2 \frac{1-exp(-2 r/r_0)}{r},
\end{equation}
used in refs.~\cite{Epele:2007ic,Epele:2008un}, will be sufficiently clarifying. The approximation consists in substituting the true wave functions by Coulomb wave functions of high $n$. For each $r_0$ a different value of $n$ will be best suited. We use the equation

\begin{equation}
\rho_n= 48 \alpha^2  n^2,
\label{size}
\end{equation}
to parametrize all expectation values in terms of $\rho_n$, where $\rho_n=r_M/r_{classical}$, $r_M$, the radius of the monopolium state,  relation which results from the calculation of the expectation value of $r$ in the $(n,0)$  Coulomb state; $\alpha$ is the electromagnetic fine structure constant $\sim \frac{1}{137}$. 

In terms of $\rho_{n_{min}}$ the mass of ground state monopolium becomes~\cite{Epele:2007ic,Epele:2008un}

\begin{equation}
M=  m\left(2-\frac{3}{4 \rho_{n_{min}}}\right).
\label{binding}
\end{equation}
Thus given a monopole mass $m$ and a monopolium ground state mass $M$ we determine $\rho_{n_{min}}$ which characterizes $n_{min}$. The binding energy of the ground state is thus $E_{n_{min}} = \frac{3 m}{4 \rho_{n{min}}}$. The excited states correspond to $n> n_{min}$. Let us label the excited states only by their principal quantum number $n$ since we are assuming in our simple Coulomb potential model that all the $l$ corresponding to an $n$ are degenerate, thus  their mass will be

\begin{equation}
M_n=  m\left(2-\frac{3}{4 \rho_n}\right),
\label{binding1}
\end{equation}
with $\rho_n =48\alpha^2 n^2$ and their binding energy $E_n = \frac{3 m}{4 \rho_n}$. The high lying deformed  ($l\ne 0$) states will have a larger magnetic moment and a larger size. Recall that in a Coulomb like potential

\begin{equation}
\frac{\langle n  l | r | n  l \rangle}{r_{Bohr}} \sim (\frac{3}{2} n^2 -\frac{1}{2}{  l}({ l}+1))  \sim  n^2,
\end{equation} 
thus the high lying states will increase their size approximately as $n^2$. We estimate their magnetic moment as,
 
\begin{equation}
{\mathcal M_n} \sim  g \, \langle n l | r | n l\rangle = g\, \rho_n \,r_{classical} =  \frac{3}{32\,\alpha^{\sfrac{3}{2}}\,E_n},
\label{magneticmoment}
\end{equation}
where we have used the Dirac quantization condition (DQC) $g \,e =\frac{1}{2}$ and $r_{classical} = \alpha_g/m$.  Looking at Eq.(\ref{magneticmoment}) it is clear that Rydberg states will have the largest magnetic moments. 

In the previous sections  we have considered the effect on coils of permanent magnetic moments, even in the case of the neutron, due to its long lifetime, but since the  monopolium states might have a shorter lifetime we have to rephrase our formulation. The two photon production of the monopolium excited states is determined by~\cite{Epele:2008un,Epele:2012jn} 

\begin{equation}
\Gamma^{2ph}_n(E) = \frac{2\,\beta_n^4\, m^3}{\alpha^2 M_n^2} \left(\frac{E_n}{m}\right)^{\sfrac{3}{2}}.
\label{twophoton}
\end{equation}
At leading order the dominant photo-production production rate is determined by the wave function at the origin and is only non zero for S ($l=0$) states. The $l\ne 0$ states will arise by de-excitation to lower $n$ levels. Analogously, at leading order the dominant decay into two photons is determined by the wave function at the origin, thus their decay rate is given by Eq.(\ref{twophoton}).  Note that the decay rate for the S states goes like $1/n^3$  \cite{Dirac:1930bga,Ore:1949te}.

Our interest here lies in the high lying deformed Rydberg states which are in high $l$ states, with vanishing wave function at the origin, that is to say that the monopole and antimonopole are far away from each other and therefore they do not annihilate. They will de-excite by spontaneous emission of photons to the lower levels where they will annihilate either as S states or to non leading order  as $l \ne 0$ state decays. In order to get an estimate for spontaneous emission as a function of $n$ we use the naive Bohr model ~\cite{Dicus:1983ry},

\begin{equation}
\Gamma^{se}_n(E) = 4096 \,\beta_n^4 \, \alpha \,m \left(\frac{E_n}{m} \right)^3.
\label{spontaneousemission}
\end{equation}
Note that this dependence goes like $1/n^6$, therefore lifetimes of these Rydberg states increase notably with $n$.

In Eqs. (\ref{twophoton}) and (\ref{spontaneousemission}) $\beta_n=1$ for Dirac coupling $g$ and $\beta_n=v_n$ for Schwinger's  $\beta$-coupling~\cite{Schwinger:1976fr},  $\alpha$ is the fine structure constant, which appears after using the DQC, $m$ for the monopole mass, $v_n= \sqrt{1-\frac{M_n^2}{E^2}}$ is the velocity of the monopolium state, which depends on the collision kinematics since $E$ is the center of mass energy of the two photon producing the state.

\subsubsection{The Dirac coupling scheme}
\label{Dirac}

\begin{figure}[htb]
\begin{center}
\hspace{-0.5cm}\includegraphics[scale= 0.95]{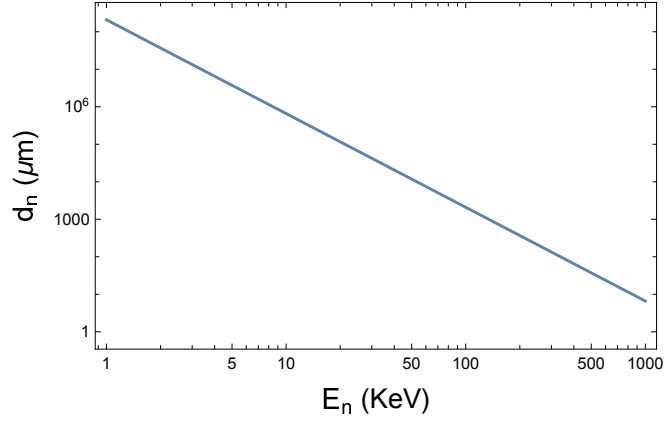} 
\end{center}
\caption{Distance in $\mu$m travelled by monopolium with binding Energy $E_n$ produced at a center of mass energy $E= 2000$ GeV before disintegrating.}
\label{distancenD}
\end{figure}

The states with the largest magnetic moment , i.e. the most deformed states will be Rydberg states with very small binding energy, $E_n << 2m$. The distance travelled by monopolium with binding energy $E_n$ for high $n$ before disintegrating is

\begin{equation}
d_n(E) = \frac{\gamma_n\,v_n} {\Gamma^{se}_n} \sim 64 \frac{v_n}{\sqrt{1-v_n^2}}\,\frac{\alpha^5}{m} \,n^6,
\label{distancetravelled}
\end{equation}
since spontaneous emission dominates for high $n$ to annihilation. We plot $d_n (E)$ in Fig.\ref{distancenD} for $E\sim 2000$ GeV as a function of $E_n$ for Rydberg states.

Despite the fact that the curve is divergent for $E_n\rightarrow 0$, i.e. $n \rightarrow \infty$, the cross section of the very excited states is so  small that with reasonable accumulated luminosities they will not be produced. We will calculate later on the highest reachable values of  $n$.

Let us  develop the formalism for states of  finite lifetime, which might be also applicable in other scenarios. For finite lifetime,
the energy stored in one coil during the lifetime of state $n$ is

\begin{equation}
{\mathcal E}_n \sim \left(\frac{12}{\pi}\right)^2 \frac{{\mathcal M}_n^2 \gamma_n\, v_n \, A^2 }{{\mathcal R}} \int_{-z{n}}^{+z_{n}}  \frac{z^2}{(z^2 + \frac{ A}{\pi})^5} dz \sim \frac{144\,{\mathcal M}_n^2 \gamma_n\, v_n }{{\mathcal R} R^3} \int_{-a_n}^{+a_n} \frac{y^2}{(y^2 + 1)^5} dy.
\label{energycoil2}
\end{equation} 
where $a_n= \frac{z_n}{R}$.

This integral can be solved analytically 

\begin{equation}
{\mathcal F} (a) = \int_{-a}^{+a} \frac{y^2}{(y^2 + 1)^5} dy = \frac{1}{192} \left(\frac{ 15 a^7 + 55 a^5 + 73 a^3 -15 a}{(1+a^2)^4} + 15 \arctan{a} \right).
\label{integral}
\end{equation}

\begin{figure}[htb]
\begin{center}
\includegraphics[scale= 0.95]{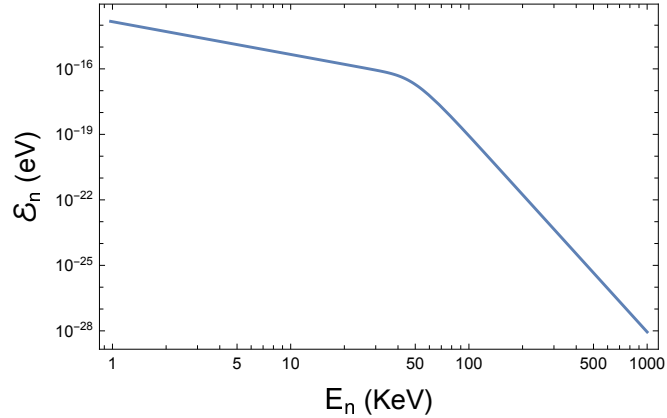} 
\end{center}
\caption{We show the energy deposited in one coil by the passing of a monopolium in state $n$ produced at a center of mass energy $E=2m$ and a monopole mass $m=1000$ GeV as a function of binding energy $E_n$ (KeV) for Rydberg states.}
\label{energycoilD}
\end{figure}

In Fig.\ref{energycoilD} we show ${\mathcal E}_n$ as a function of the binding energy  for $\mathcal R = \frac{2\pi R {\varrho}}{\pi r_s^2}$ with the radius of the coils $R=10$ mm, the radius of the section of the coil $r_s=1$ mm and the resistivity ${\varrho}=10^{-8}$ Ohm$\cdot$m.

We next assume an accumulated  luminosity ${\mathcal L}$ and calculate the effect of all the states produced during this running. The cross section for monopolium production by photon fusion, which is the dominating process at large energies, is given for each bound state  $n$ by \cite{Epele:2007ic,Epele:2008un}

\begin{equation}
\sigma(2 \gamma \rightarrow  {\mathfrak M}_n)(E) = \frac{4 \pi}{E^2} \frac{M_n^2\, \Gamma^{2ph}_n(E) \Gamma_{M_n}}{(E^2-M_n^2)^2 + M_n^2 \Gamma_{M_n}^2},
\label{sigma}
\end{equation}
where $M_n$ is the mass of the $n$th state, $\Gamma^{2ph}_n(E)$ characterizes the production rate and $\Gamma_{M_n}$ arises from the softening of the delta function $\delta(E^2-M_n^2)$. In the case of LHC the photons originate from composite particles (protons) and this expression has to be convoluted with their structure. Also an equivalent photon approximation for the photon spectrum of the intermediate constituents has to be used. This process has been described in detail for the calculation of the ground state monopolium cross section by proton-proton collisions at LHC in ref. \cite{Epele:2007ic,Epele:2008un}. We use the latter calculation to get an estimate for the present calculations.  Let $\sigma_{n_{min}}(E)$ be the cross section calculated for the ground state then the cross section for the $n$ state will be approximately given by

\begin{equation}
\sigma_n(E) \sim \frac{\sigma(2 \gamma \rightarrow  {\mathfrak M}_n)(E)} {\sigma(2 \gamma \rightarrow  {\mathfrak M})(E)} \sigma_{n_{min}}(E).
\end{equation}
Ground state monopolium for a binding energy of $\frac{2 m}{15}$ leads to $n_{min} \sim 47$ and an estimate for $\sigma_{n_{min}}(E)\sim 50$ fb at the $14$ TeV LHC in the region of interest, $E\sim 2m$, in the case of Dirac coupling with a monopole mass $m \sim 1000$ GeV~ \cite{Epele:2007ic,Epele:2008un}.

Now for an accumulated luminosity ${\mathcal L}$ the number of states created for a center of mass energy $E$ is $\sigma_n(E){\mathcal L}$. The maximum excited state for an energy $E$ will be obtained by solving for $n$ the following equation

\begin{equation}
\sigma_n(E){ \mathcal L} = 1
\end{equation}

\begin{figure}[htb]
\begin{center}
\includegraphics[scale= 1.0]{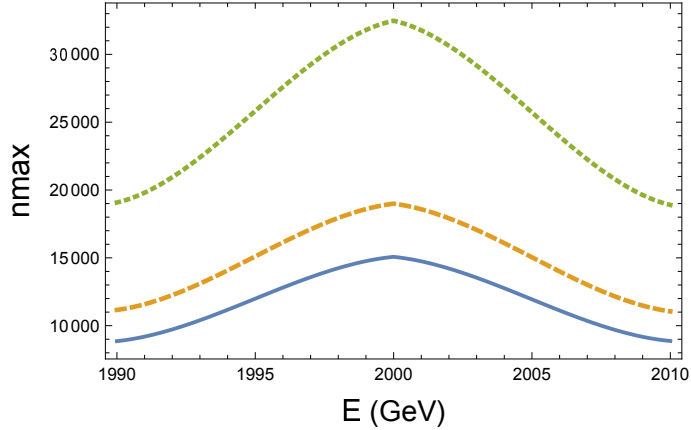} 
\end{center}
\caption{We show $n_{max}$ in the region of interest around   $E=2m$ for luminosities  of $1000$ (solid), $2000$ (dashed) and $10000$ (dotted) fb$^{-1}$,   a monopole mass $m=1000$ GeV as a function of center of mass energy $E$ (GeV).}
\label{nmaxx}
\end{figure}

The outcome of the calculation, for accumulated luminosities of $1000$, $2000$ and $10000$ fb$^{-1}$, leads to $n_{max}$ as a function of E as plotted in Fig. \ref{nmaxx}. 

If we now sum the contribution of all states to a single coil deposited energy we get

\begin{equation}
{\mathcal E}_{total}(E) \sim \sum_{nmin}^{nmax} {\mathcal E}_n(E) n^2 {\mathcal L} \sigma_n(E),
\end{equation}
where $n^2$ is associated to the degeneracy of the states.

\begin{figure}[htb]
\begin{center}
\includegraphics[scale= 1.0]{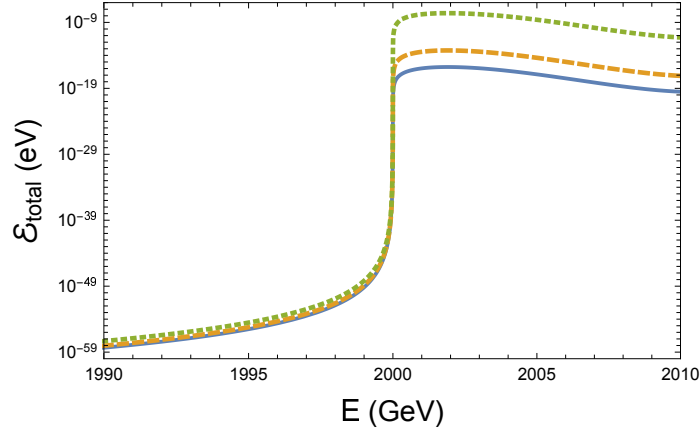} 
\end{center}
\caption{We show the energy deposited  in a single coil ${\mathcal E}_{total}(E)$ by the passing of  all monopolia produced by accumulated luminosities at LHC of  $1000$ (solid), $2000$  (dashed) and $10000$ (dotted) fb$^{-1}$, as a function of center of mass energy $E$ (for $m=1000$ GeV) .}
\label{energyDSEtotal}
\end{figure}

The results for accumulated luminosities of $1000$, $2000$ and $10000$ fb$^{-1}$ are shown in Fig. \ref{energyDSEtotal}.
${\mathcal E}_{total}(E)$ saturates close to $E \sim 2m$, where the highest Rydberg states contribute and drops slowly above due to the cross section drop away from the resonances. The signal for monopolia to be expected is a sudden jump in the deposited energy, as shown in Fig.\ref{energyDSEtotal}, followed by disintegration into photons a few micrometers away, that will be captured by the LHC photon detectors. The current signal is small $\sim \mu eV$ but measurable. Moreover, the excited states produce two types of very distinct photons, the ones originating from annihilation~\cite{Epele:2012jn,Barrie:2016wxf,Fanchiotti:2017nkk,Barrie:2021rqa}  and the photons produced by cascading  to lower states. The latter arising from Rydberg states will have much lower energy than the annihilation photons. With the present parameters the latter  have energies energies of KeV while the former of GeV. Lastly, another  way to see these  monopolium states would be by the deformation of the proton beam~\cite{Vento:2019auh} after having seen a jump in the coil detector.

To end this section it must be stressed that the luminosity requirement given the small cross section of the high Rydberg states sets a huge limitation on the coil energy and the distance before disintegration. We will come back to this point in Section \ref{cosmic}.

\subsubsection{The Schwinger coupling scheme}
\label{Schwinger}

\begin{figure}[htb]
\begin{center}
\includegraphics[scale= 0.95]{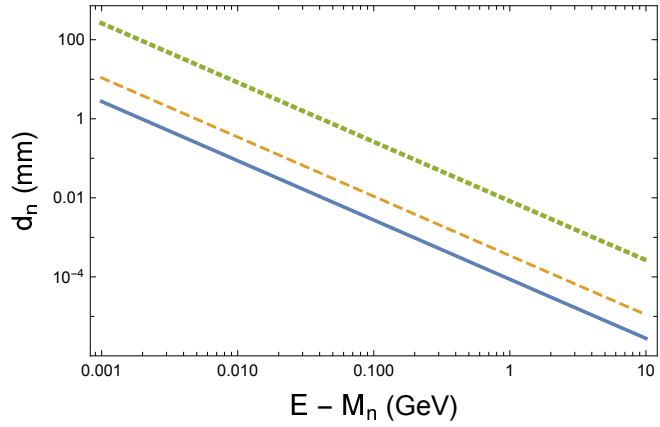} 
\end{center}
\caption{We show the distance travelled by the highest Rydberg monopolia states produced by LHC luminosities of $1000$ (solid), $2000$ (dashed) and $10000$ (dotted) fb$^{-1}$ as a function of the energy away from threshold.}
\label{distancenbetaSE}
\end{figure}

The velocity in Eq.(\ref{spontaneousemission}) in Schwinger coupling makes the lifetime very short at threshold and therefore we expect some interesting physics around the threshold region. Let us calculate the value  of $d_n$ at threshold

\begin{equation}
\lim_{E \rightarrow M_n} d_n (E) \sim \frac{m^2}{4096\, \alpha \,E_n^3} \left(\frac{M_n}{2 (E-M_n)}\right)^{\sfrac{3}{2}}.
\end{equation}
The distance travelled by monopolia diverges at threshold and we expect physics outside the beam region. Unluckily the production cross section tends to zero in that region. Thus close to threshold we expect a small number of slow moving excited monopolia traveling for long distances as shown in Fig. \ref{distancenbetaSE}. Note that we are limiting the results of the figure to LHC luminosities. For cosmologic monopolia the distances will be shown to be much greater.

\begin{figure}[htb]
\begin{center}
\includegraphics[scale= 0.95]{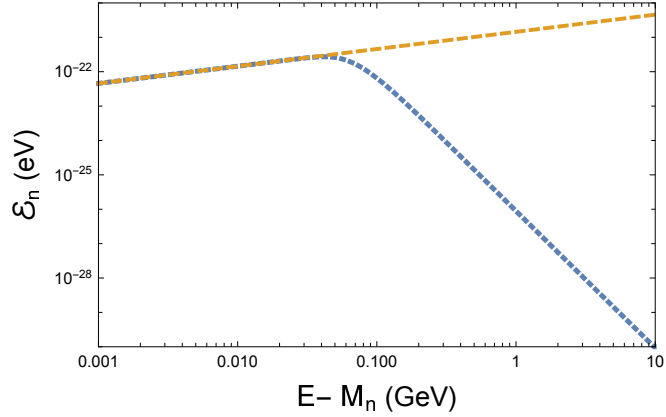} 
\end{center}
\caption{We show the energy deposited in one coil by a monopole state  of binding energy $E_n= 100$ MeV as a function of $E - M_n$. The dashed line shows the limiting value Eq.(\ref{thresholdlimit}) while the dotted line shows the exact calculation Eq.(\ref{thresholdexact})}.
\label{energycoilbetaSE}
\end{figure}

Let  us calculate now the size of the deposited energy. Using Eqs.(\ref{energycoil2}) and (\ref{integral}) we get

\begin{equation}
{\mathcal E}_n = 1.650 \, 10^{-29} \frac{{\mathcal M}_n^2 \gamma_n \, v_n r_s^2 {\mathcal F} (a_n)}{ R^4\, \varrho} \;\; \mbox{eV},
\label{thresholdexact}
\end{equation}
where $r_s$ and $R$ are in mm, $\mathcal{M}_n \sim \frac{3}{32 \alpha^{\sfrac{3}{2}} E_n}$ in GeV$^{-1}$ and $\varrho$ in Ohm$\cdot$m. At threshold $a_n \sim \frac{d_n}{2\,R} \rightarrow  \infty$, thus ${\mathcal F}(a_n) \rightarrow \frac{5 \pi}{128}$, and therefore

\begin{equation}
\lim_{E\rightarrow M_n}  {{\mathcal E}_n}\sim 1.450\,10^{-31} \frac{ r_s^2 }{R^4 \alpha^3 \varrho} {\mathcal F}(a_n) \frac{1}{E_n^2}\left(\frac{E- M_n}{M_n}\right)^{\sfrac{1}{2}} \sim 0.6471\,10^{-21} \frac{1}{E_n^2}\left(\frac{E- M_n}{E}\right)^{\sfrac{1}{2}} \;\;  \mbox{eV},
\label{thresholdlimit}
\end{equation}
where in the last expression we have used $R =10$ mm, $r_s=1$ mm, $\varrho=10^{-8}$ Ohm$\cdot$m and $\alpha=\frac{1}{137}$. We note that the deposited energy tends to zero at threshold. However if we look carefully, as shown in Fig. \ref{energycoilbetaSE}, we see that there is a maximum of the deposited energy close to threshold.

Eq.(\ref{thresholdexact}) is divergent if we let $n$ run to infinity. However, as we saw in the Dirac case the higher lying states are not produced once we fixed the accumulated luminosity leading to a finite $n_{max}$ in the expression for the total energy,

\begin{equation}
{\mathcal E}_{total}(E) \sim \sum_{nmin}^{nmax} {\mathcal E}_n (E)n^2 {\mathcal L} \sigma_n(E).
\label{Etotal}
\end{equation}

\begin{figure}[htb]
\begin{center}
\includegraphics[scale= 0.95]{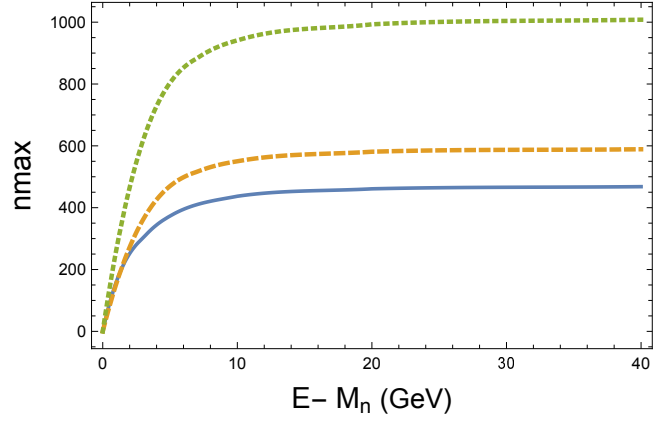} 
\end{center}
\caption{We show the variation of $n_{max}$  way from threshold for three accumulated luminosities of $1000$ (solid) , $2000$ (dashed) and $10000$ (dotted) fb$^{-1}$.} 
\label{nmaxxS}
\end{figure}

We follow the same procedure as before to obtain $n_{max}$. By using  the cross section for $E$ slightly above $M_n$  and recalling that in this case the ground state monopolium cross section for Schwinger coupling is much smaller ($\sim 1$fb). We show in Fig. \ref{nmaxxS} three cases  of $n_{max} $ as a function of the difference in energy between threshold and measured. The cases correspond to ${\mathcal L}= 1000, 2000$ and $10000$ fb$^{-1}$. We see that $n_{max}$ stabilizes beyond 10 $GeV$ away from threshold around $\sim 460, 580, 1000$. Since $n_{max}$ describes the Rydberg states it is clear that the higher the luminosity the higher Rydberg states will be excited and therefore the larger will be the magnetic moment and the distance travelled by the monopolia before decaying. Compared with the Dirac case, since the coupling  now is much weaker close to threshold, the values of $n_{max}$ are smaller and therefore the deposited energy is smaller despite the fact that the distance travelled is larger.

\begin{figure}[htb]
\begin{center}
\includegraphics[scale= 0.95]{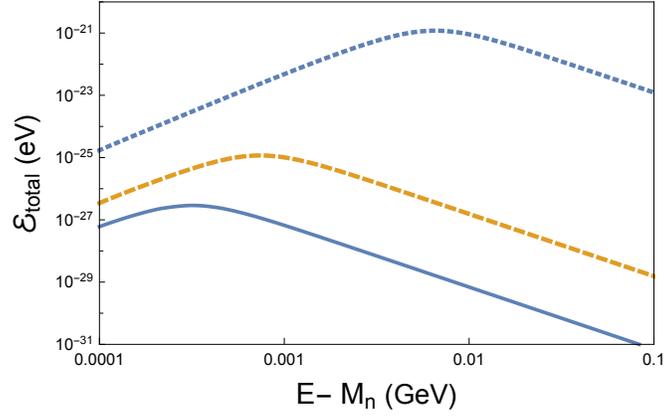} 
\end{center}
\caption{We show the energy deposited  in a single coil by the monopole excited states for accumulated luminosities of $1000$ (solid), $2000$ (dashed) and $10000$ (dotted) fb$^{-1}$.}
\label{energycoilbetaSE3}
\end{figure}

Let us next show the total deposited energy of Eq.(\ref{Etotal}) in Fig. \ref{energycoilbetaSE3} for the same three luminosities.
The values are very small due to the relative  low values of $n_{max}$. In this case the distances travelled by monopolia for the three luminosities are  shown in  Fig.\ref{distancenbetaSE} and we notice that the distances where the values of $n_{max}$ are maximum  are  small, thus the use of a solenoid of a large number of coils, as discussed in Section \ref{faraday}, is out of question. The deposited  energy is therefore too small to be detectable by actual means in this case. However, one might still look for cascading and decay photons. The former in this case will have energies of MeV.

\begin{figure}[htb]
\begin{center}
\includegraphics[scale= 0.95]{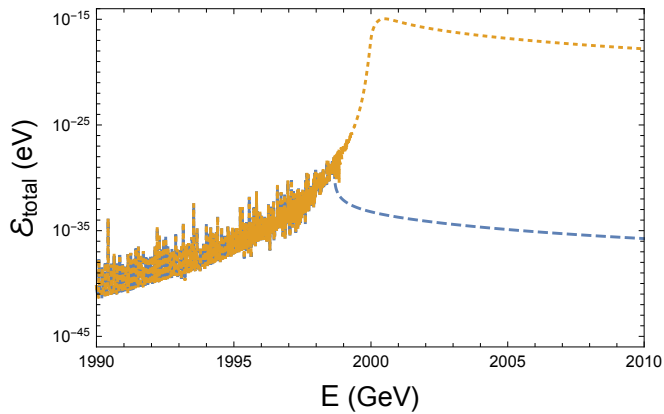} 
\end{center}
\caption{We show the energy deposited in a single coil by the monopole excited states for an accumulated luminosity of $1000$ (solid) , $2000$ (dashed) and $10000$ (dotted) fb$^{-1}$ as we sweep across the monopole-antimonopole thresholds and above $2m$.}
 \label{energycoilbetaSEtotal}
\end{figure}

If we sweep across $2m$ we see the characteristic feature of Schwinger coupling: peaks at the threshold values and the fall off away from the thresholds as shown in Fig. \ref{energycoilbetaSEtotal}. This characteristic feature, for higher values of the deposited energy, will occur in those scenarios with very large values of $n_{max}$.

\subsubsection{ Summary of accelerator experiments}
\label{summary}
Let us summarize our findings in both subsections of collider experiments. The experimental scheme discussed thus far is based on two crucial properties of monopolia, namely their large magnetic moments in deformed states and their decay into photons.  An ideal experimental scenario would be to have a large flux of Rydberg states of monopolium  with very high $n$, this would produce a large current in the coils and large decay distances.  Unluckily this does not  happen in collider experiments due to the small cross sections of the high lying Rydberg states and the limited luminosity. However, this study has deepen our understanding of the properties of monopolia. We have done so by discussing  two schemes under the name of Dirac and Schwinger coupling scenarios. The Dirac scenario is characterized by a strong coupling at threshold and therefore the distance travelled by monopolia before decaying is small, however it leads to relatively  high $n$ excited states. The outcome of our calculation shows that in this scenario we are on the verge of detection at colliders  if one is able to measure  deposited energies at the coils of $\mu eV$. Monopolia in this case  decay inside the beam region with decay distances of $\mu m$. Thus the signal for monopolia in colliders is a jump in the deposited energy followed by a decay into photons.The photons are of two types: cascading photons, which can be of relatively low energy, and annihilation photons which are of very high energy. The Schwinger coupling scenario on the other hand might lead to larger decay  distances, mm, close to threshold but the deposited energies are small and difficult to detect (feV). This features have to do with the fact that the excited  Rydberg states are not very high $n$ states due to their low cross sections near threshold. Away from threshold, i.e. were $v_n \sim 1$ both Schwinger and Dirac schemes are identical. In the Schwinger scenario the coil scheme is not suitable for collider experiments. One has then to resort only to the study of photon decays as done sometime ago \cite{Epele:2012jn,Barrie:2016wxf}, but in the case of excited states one has to look not only for photons of very high energies, GeV, arising from disintegration, but also at  cascading photons with lower energies, KeV to MeV, for the highest Rydberg states.
If by other means one could create very high $n$ Rydberg states they would have have long decay paths and a  combination of solenoids with the MAPP detector~\cite{Mitsou:2020hmt} would be an ideal experimental tool for their detection and characterization. We see next that this might happen in astronomical observations. To conclude,  Figs. \ref{energyDSEtotal} and  \ref{energycoilbetaSEtotal} summarize the different behavior of monopolia in the two coupling schemes: the Dirac scheme shows a sudden jump about the zero binding energy point ($E=2m$), while the Schwinger scheme shows peaks for each threshold ($E=M_n$) accumulating as we approach the zero binding energy point.

\subsection{Cosmic monopolium}
\label{cosmic}
Let us pursue now the observation of cosmological  monopolium states. Our setup is particularly suited for heavy monopolia  originating from GUT  or Kaluza-Klein theories, which cannot be produced in accelerators. However, we will extend their detection analysis also to light monopolia because the very high $n$  Rydberg states of light monopolium before dissociation in the Earth's magnetic field have large lifetimes as we have already seen. The discussion related to the effect of the induced current on the particle velocity can be easily generalized to cosmic monopolium. The huge mass of the monopolium states leads to small accelerations and therefore the effect on the traveling particle velocity is here also negligible.

\subsubsection{Light monopolium}
\begin{figure}[htb]
\begin{center}
\includegraphics[scale= 0.8]{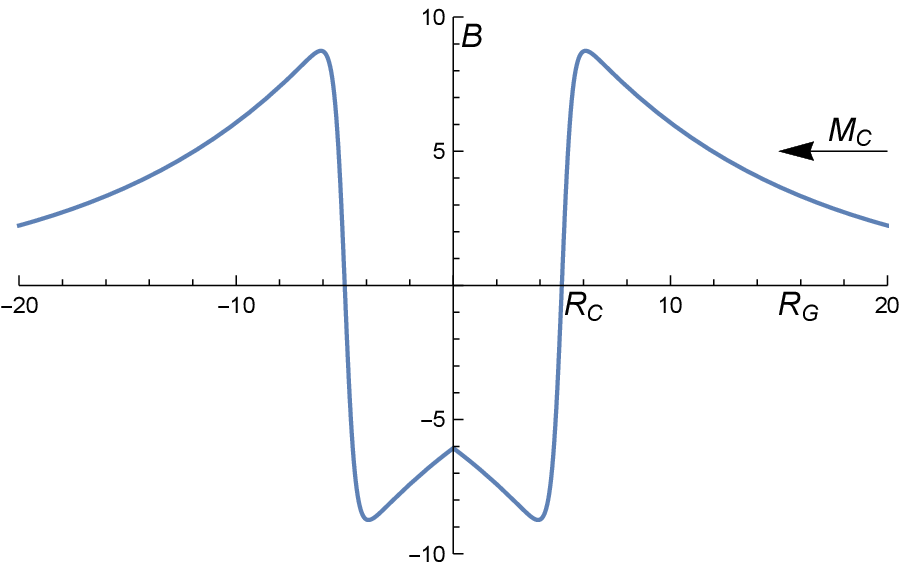} \hspace{1cm} \includegraphics[scale= 0.8]{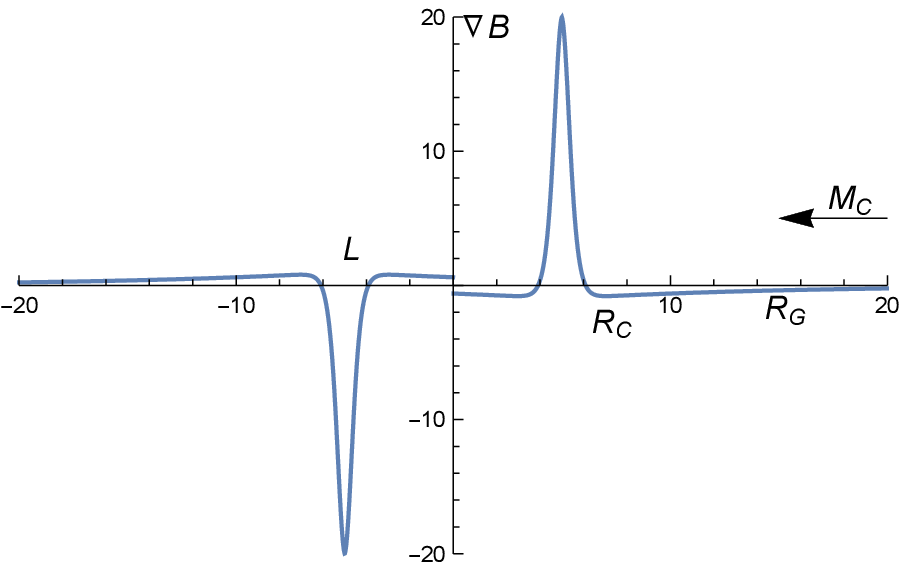} 
\end{center}
\caption{We show left our one dimensional model for the magnetic field of a galaxy and right its gradient. $M_C$ represents a monopolium cloud traveling towards the center of the galaxy.}
 \label{galaxy}
\end{figure}

In the case of light monopolia ($m \sim 1000$ GeV) we have to generalize the calculation of their velocity in a galaxy and find a reasonable flux bound. Monopolia will interact with the cosmological magnetic field through their magnetic moment, thus the force acting on them will be given by

\begin{equation}
\vec{F} = - \vec{{\mathcal M}} \cdot \vec{\nabla} \vec{B},
\end{equation}
where ${\mathcal M}\sim g d $ is the magnetic moment, $d$ represents the distance between the poles and $\vec{B}$ the cosmological magnetic field. \begin{eqnarray}
B(R)  = &\;\,B_0 \;\tanh{\left(\frac{R-R_C}{R_1}\right)} \exp{\left(-\left|\frac{R-R_C}{R_0}\right|\right)}  \hspace{1.5cm}&{\mbox  for} \; R > 0, 
\nonumber\\
  = &- B_0 \;\tanh{\left(\frac{R+R_C}{R_1}\right)} \exp{\left(-\left|\frac{R+R_C}{R_0}\right|\right)} \hspace{1.5cm}&{\mbox  for} \; R < 0 .  \nonumber \\
  & 
  \label{galaxyEq}
 \end{eqnarray}
In the left hand side of Fig. \ref{galaxy} the model for the  magnetic field of a galaxy described in Eq. \ref{galaxyEq} is shown~\cite{Vento:2020vsq} and in the right hand side  its gradient along the model galaxy.

\begin{figure}[htb]
\begin{center}
\includegraphics[scale= 0.95]{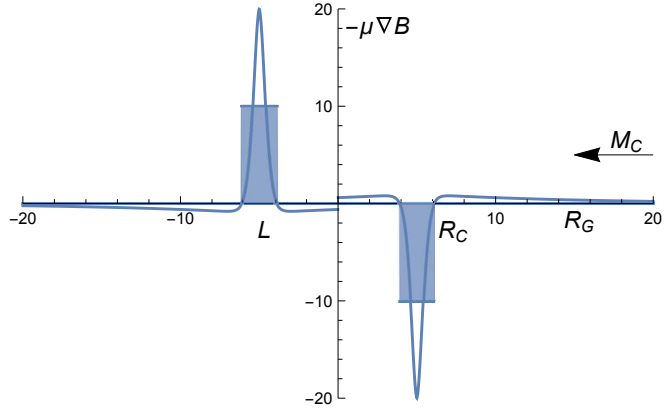} 
\end{center}
\caption{The simplified force field used in the calculation of the velocity is in grey over imposed on the real force field. $L$ is the width of the simplified force field.}
 \label{force}
\end{figure}

The mathematical structure of the field makes the analysis of the velocity of monopolium complicated. In order to get an estimate we substitute the  force field by two simplified forces of adequate strength as shown in Fig. \ref{force} compared to the actual force field. With this constant force field the velocity becomes,  after crossing the two accelerating structures, and using relativistic kinematics

\begin{equation}
v_M \sim \sqrt{1- \frac{1}{(\frac{g d B_0 L}{2 M R_1}+1)^2}} \sim \sqrt{1- \frac{1}{(0.32\, 10^{-26} \,d +1)^2}}\,.
\label{velocity}
\end{equation}
The last expression has been calculated for a high $n$ Rydberg state, $M \sim 2000$ GeV, and for conventional values of the galaxy parameters  $B_0 = 10 \mu$G $\sim1.95 \;10^{-25} $ GeV$^2$, $R_1= 0.5 $ Kpc, where the width of the force field $L \sim 2.2$ Kpc, with $d$ measured in fm.  Despite the huge sizes of Rydberg states ($d \sim 10^9$ fm $\sim 10^{-3}$ mm) the small prefactor in the formula cannot be overcome and the velocity originating from the gravitational magnetic fields will be much smaller than the galactic gravitational velocity $v_M \sim 10^{-3} $. Moreover, latter  is mass independent~\cite{Preskill:1984gd}.

The  Rydberg states with long lifetimes and large magnetic moments are lightly bound monopole-antimonopole pairs with binding energies of the order of eVs and  the small Parker bound can be obviated and detection is possible~\cite{Dicus:1983ry}. The Parker bound is based on the monopoles taking energy from the intergalactic magnetic field. However, monopolia below the dissociation energy do not take energy from the magnetic field since they are almost neutral, i.e., magnetic moment interactions are much weaker than magnetic charge interactions.
However, they might be dissociated by the much stronger Earth's magnetic field leading to free monopole-antimonopole pairs which might annihilate or ionize metal rods depending on its production kinematics. Monopolia below the dissociation limit survive and those close to the dissociation limit have long lifetimes, of the order of  minutes for Dirac coupling, and much greater lifetimes for Schwinger coupling close to threshold. Thus in this case  their paths before annihilation will be very long and we can use long solenoids with many coils increasing detection efficiency enormously.

Let us calculate the maximum $n$ before dissociation by equating the Energy of the Rydberg state, $E_n$ with the dissociation energy \cite{Dicus:1983ry} 

\begin{equation}
E_n \sim g B_E r_n.
\label{dissociation}
\end{equation}
Using the $E_n = \frac{3 m}{4 \rho_n} = \frac{3 g^2}{4 r_n}$ and the Earth's magnetic field $B_E \sim 1$ G $\sim 1.95 \,10^{-20}$ GeV$^2$, we get for the maximum allowed size $r_n \lesssim \sqrt{\frac{3}{8 e B_E}} \sim 3\, 10^{-3}$ mm. Using now $r_n \sim \rho_n r_{classical}$ we get  the $n$ for dissociation $n \sim \sqrt{\frac{r_n m}{12 \alpha}}$, where m is the monopole mass. For our light monopole $m \sim 10^3$ GeV, $n \lesssim 10^7$ and a binding energy $E_n  \gtrsim 1 eV$

\begin{figure}[htb]
\begin{center}
\includegraphics[scale= 0.9]{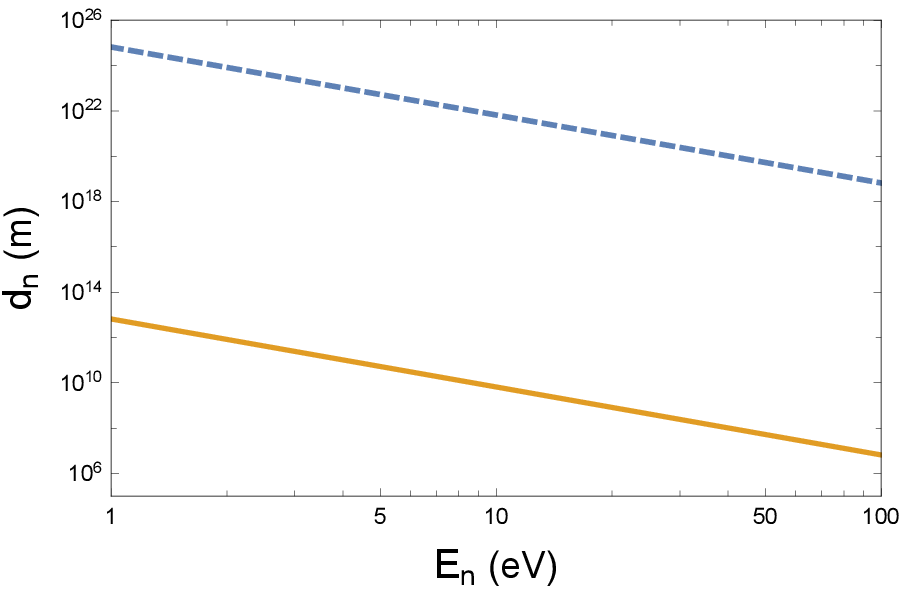} \hspace{0.5cm}\includegraphics[scale= 0.9]{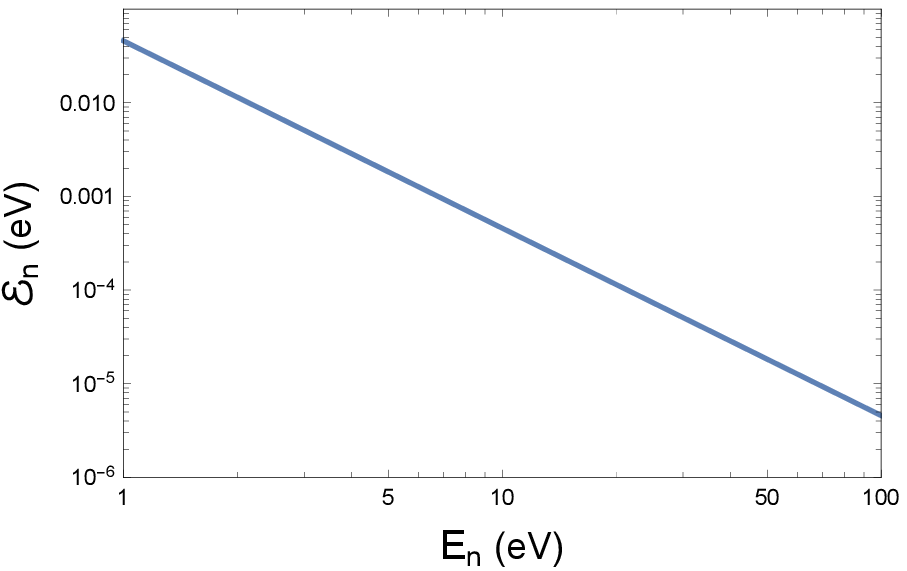} 
\end{center}
\caption{Left: The decay distance travelled by monopolia of high $n$: Dirac coupling (solid) and Schwinger coupling (dashed), as a function of binding energy. Right:The deposited energy on a solenoid of the characteristics explained in the text as a function of the binding energy. For fixed binding energy the deposited energy is the same for both Dirac and Schwinger couplings.}
\label{CosmicMonopolium}
\end{figure}

Let us assume that we have one of those monopolia of $d\sim 10^{-3}$mm reaching a solenoid of $10$ m long with $N \sim 10^5$ no tight coils made of a good conductor $\varrho \sim 10^{-8}$ Ohm$\cdot$m, with $R\sim 10$ mm and $r_s \sim 1$ mm. At a speed  $v_M \sim 10^{-3}$ the solenoid will be transversed since the decay distances are very large, as shown in Fig. \ref{CosmicMonopolium} . The solenoid will get a deposited energy for the highest Rydberg states of $10^{-6} - 10^{-1}$ eV per monopolia as shown in Fig. \ref{CosmicMonopolium}.  This  experimental setup is suited to detect single events in the line with the original monopole detectors \cite{Cabrera:1982gz,Milton:2006cp,MoEDAL:2014ttp,Acharya:2014nyr,Patrizii:2015uea}. A more sophisticated MoEDAL type detector, implemented with solenoids and MAPP \cite{MoEDAL:2014ttp,Mitsou:2020hmt} would be an ideal detector. The Earth's magnetic field might break up some high lying monopolia Rydberg states and  pairs of monopole-antimonopoles would be produced. These  could be detected by MoEDAL's traditional design. If the pair annihilates into photons MAPP could detect these photons. However, if the Rydberg states are sufficiently bound and not broken by the Earth's magnetic field, they produce a sudden jump of the energy deposited in the solenoids, and  MAPP will detect the low energy photons from the cascading Rydberg states. Note that given their long lifetimes annihilation, high energy photons, occurs beyond MAPP.  The  cascading photons between the upper Rydberg states will be of  low energy ($1 -100$ eV)  and easy to detect. To determine the mass, since the velocity of monopolium is known, one has to resort to kinematics locating an adequate target in the MAPP detector, and having monopolia interact with charged particles which are easy to detect. Once their mass is known a spectroscopic analysis will determine the dynamics and the mass of the monopole.

\subsubsection{Heavy monopolium}

This scenario is thought for extremely heavy monopoles and therefore very heavy monopolia. For GUT monopoles the Parker bound can be overcome when monopole-antimonopole bind. In this case, by equating the binding energy to the dissociation energy as above but using a mass for the monopole of $10^{16}$ GeV, we get  a limiting value for monopolia of $n \lesssim 5\, 10^{13}$ which corresponds to  binding energies $E_n \gtrsim 1$ eV and sizes of $d  \lesssim10^{-3}$  mm as before. The velocity of the monopolia, in this case, is also the intergalactic gravitational  velocity $v_M \sim 10^{-3}$ which is independent of mass. The crucial ingredients of our previous calculation, binding energy and size,  remain the same and the analysis follows previous calculation for the very high Rydberg states leading to results similar to those shown in  Fig.\ref{CosmicMonopolium} for the deposited energy. The decay distance  is much larger since the ratio $(E_n/m)^3$ entering the spontaneous emission lifetime, Eq. (\ref{spontaneousemission}), is much smaller. Despite the difference the decay distances are so large that for both light and GUT monopolia their decay  will occur far outside the apparatus. The  low energy cascading photons are of similar energy. Thus we see no new physics arising for the GUT monopoles except their mass which should be measured by a kinematical detector within our setup.

The idea behind Kaluza-Klein (KK) theories is that the world has more than three spatial dimensions and some of them are curled up to form a circle so small as to be unobservable  \cite{Kaluza:1921tu,Klein:1926tv}. KK theories have been the subject of revived interest in recent years since many standard model extensions in extra dimensions yield KK theories. KK theories contain a very rich topological structure, which includes very heavy monopoles whose mass is around the Planck mass  \cite{Gross:1983hb,Sorkin:1983ns}. Most important  they also contain other soliton solutions in different topological sectors. In particular  the dipole, which has the quantum numbers of a monopole-antimonopole bound state. As we have seen above, in conventional gauge theories monopolium has vacuum quantum numbers and annihilates. However, in KK theories monopolium does not belong to the topological sector of the vacuum and therefore it is classically stable  \cite{Gross:1983hb}.  In realistic scenarios the dipole does not have vacuum quantum numbers~\cite{Newman:1963yy,Hawking:1976jb,Gibbons:1979xm,Vento:2020vsq} and its structure is described in terms of some parameter $d$ (distance between center of the poles), which approaches the monopole-antimonopole structure as  $d$ becomes large. The dipole can thus be interpreted as a bound state with a mass smaller than twice the monopole mass. The parameter $d$ describes the magnetic moment of the dipole. The ground state (smallest mass) dipole, i.e. when $d\rightarrow 0$, has vanishingly small dipole moment and therefore it is electromagnetically neutral and thus has gravitational interaction only. When $d > 0$ we can ascribe these states to excited states of monopolium and in the large $d$ limit we are discussing, the equivalent of Rydberg states. since those states are very much like an elongated  monopole-antimonopole structure separated by a distance $d$ the magnetic moment is 

\begin{equation}
{ \cal M} \sim g\, d.
\end{equation}

Once this similarity is exploited, again the parameters that play a role in the calculation are  the monopole mass $m \sim M_{Plack}\sim\sqrt{\frac{\hbar \,c}{G}} \sim 10^{19} GeV$ where $G$ is the gravitational constant, the binding energy, $E_d = 2M - M_d$ and the mass independent gravitational velocity $v_d\sim 10^{-3}$. In this case again binding energies and sizes will be determined by avoiding dissociation.  Let us use for our calculation the mass formula for the dipole  of the Gross-Perry model \cite{Gross:1983hb,Vento:2020vsq}

\begin{equation}
M_d = 2 \left(m-\frac{M_d^2}{\sqrt{4 M_2^2 + \frac{d^2}{G^2}}}\right),
\end{equation}
where $G$ is Newton's constant $1.32 \, 10^{-39}$ fm/GeV, $m \sim \frac{M_{Planck}}{4 \sqrt{\alpha}}$ the monopole mass, with $M_{Planck} \sim 1.22 \, 10^{19}$ GeV, $M_d$ the monopolia mass and $d$ is given in fm and the masses in GeV. For large $d$ we obtain  the binding energy approximation for the Rydberg states

\begin{equation}
E \sim \frac{8 G m^2}{d}.
\end{equation}
Equating this energy to the dissociation energy $E = g B_E d \sim \frac{g B_E d}{0.197 GeV\cdot fm} $, where $B_E$ is measured in GeV$^2$ and $d$ in fm, we get the maximum monopolium size in the Earth at $ d = \sqrt{\frac{1.58 GeV \cdot fm G m^2}{g B_E}}$ which turns out to be $d \sim 10^{-2}$ mm with a binding energy is $E \sim 1 $ eV.  Size and binding energy are very similar to the ones above. The physics is going to be the same, and therefore the experimental setup is also suitable for this monopole dynmics. Only a kinematic detector will distinguish between the three monopoles studied. The remaining detectors determine the presence of monopoles or monopolia but are not able to distinguish  which type of monopoles are being detected. The fact that they KK monopolia are stable does not make any difference for detection purposes because the decay distance in gauge theories is very large, thus from the point of view of our setups all monopolia look stable. We foresee in this case two mechanisms of production given their huge mass, either a primordial mechanism which took place before inflation~\cite{Vento:2020vsq} or a cataclysmic collision of monopolium clouds after inflation. The former will produce mostly ground state or low excited monopolia with small values of $d$, very difficult to detect with the present setup. The  latter will produce highly excited states with large $d$, but very few in number. Similarly to our previous discussion we need detectors which look at extraordinary events and the presented setup, i.e. MoEDAL monopole detector together with  by solenoids and  MAPP  supplemented by some kinematic detector, is well suited~\cite{MoEDAL:2014ttp,Mitsou:2020hmt} .

\section{Concluding remarks}
\label{conclusion}

Monopolium, a bound state of monopole-antimonopole, has no magnetic charge and in its ground state no magnetic moment being very difficult to detect directly. In conventional gauge theories it has the quantum numbers of the vacuum and annihilates into photons, and these disintegrations have been intensively studied \cite{Epele:2007ic,Epele:2008un,Epele:2012jn,Barrie:2016wxf,Fanchiotti:2017nkk,Barrie:2021rqa}. In Kaluza-Klein theories monopolium is classically stable and therefore very long lived \cite{Gross:1983hb,Vento:2020vsq} but its detection problems still persist since it only interacts gravitationally in vacuum. 

 What happens with excited monopolium states? This paper deals with the study of the excited states of monopolium. Excited states have permanent magnetic moments and their interaction with magnetic and electric fields is known.  We have used a theoretical analysis of magnetic dipole detection to learn about excited monopolium states. Analyzing the detailed characteristics of excited monopoliun states we have discovered many properties, in particular of the high $n$ Rydberg states.
 
 The first part of the analysis deals with a theoretical description of the interaction of  magnetic moment with coils and solenoids both for conductors and superconductors.  Given that the difference between the two is not large and that our aim is to learn about the behavior of Rydberg states more than to perform a detailed experimental analysis of detection, we have continued only with conductors. We have applied our analysis to the neutron, which for the purposes of the analysis has been treated as a stable particle, describing a possible experiment to measure the magnetic moment of the neutron. 
 
 We continue with the application to the detection of excited monopolium in accelerators which requires a reformulation of the previous analysis to adapt to states with finite lifetimes which affect the behavior of the coils. We have done the calculation for the two most used schemes, the Dirac scheme and the Schwinger or $\beta$ scheme.  We have seen that, due to the low production cross section of excited monopolium, the detection via the Faraday-Lenz analysis is very difficult. There might be a slight chance of detection in Dirac coupling, if one is able to measure relative low coil energies, but no chance in Schwinger coupling. The Dirac coupling has the inconvenience of very short decay distances which implies that the decay will occur inside the beam. In both cases the excited Rydberg states will de-excite via  photons of energies in the KeV to MeV range, which compared to the decaying photons which are at the level of hundreds of GeV, are clear signals of excited monopolia. Thus besides the deposited energy,  the detection of these cascading photons together with the energetic annihilating ones, are clear signatures of excited monopolia.

Very heavy monopoles cannot be produced in accelerators. However, nature might provide us with them in the Cosmos. Since bound states do not absorb energy from the galactic magnetic fields, monopolia can therefore overcome the Parker bound. When these excited monopolia reach the Earth, the Earth's magnetic field might break them giving rise to monopole-antimonopole pairs, which might not annihilate since the excited states are very large, $10^{-3}-10^{-2}$ mm. Typical magnetic monopole detectors can be successful in seeing them~\cite{Cabrera:1982gz,Milton:2006cp,MoEDAL:2014ttp,Acharya:2014nyr,Patrizii:2015uea}. However, for larger binding energies the monopolium Rydberg states remain and they are ideal candidates for the setup presented here.  These states, as we have shown,  have large decay distances and therefore appear as stable for the detectors. Thus, we can use large solenoids which increase the deposited energy. These states emit extremely soft cascading photons $ \sim eV$. All these properties point that a Cosmological detector of  MoEDAL type, supplemented by MAPP and Faraday-Lenz solenoids could be ideal to detect monopoles and/or excited monopolia.

\section*{Acknowledgement}
Vicente Vento would like to thank Nick Mavromatos, Vassia Mitsou and Jim Pinfold for useful conversations.
This work was supported in part  by the MICINN and UE Feder under contract FPA2016-77177-C2-1-P.

\end{document}